\documentclass[10pt]{iopart}
\usepackage{ccaption}
\usepackage{graphicx}
\usepackage{morefloats}
\usepackage{setstack}
\usepackage{color}
\usepackage{epsf,epsfig,graphics}
\def\etal{{\it et al~}}

\begin{document}

\title[W$^{13+}$ - W$^{1+}$ DR rate coefficients]{Dielectronic recombination of 
lanthanide and low ionization state tungsten ions: W$^{13+}$ - W$^{1+}$}

\author{S. P. Preval$^{1,2}$}
\author{N. R. Badnell$^{1}$}
\author{M. G. O'Mullane$^{1}$}

\address{$^{1}$Department of Physics, University of Strathclyde, Glasgow G4 0NG, United Kingdom}
\address{$^{2}$Department of Physics and Astronomy, University of Leicester, University Road, Leicester, LE1 7RH, United Kingdom}

\ead{spp11@leicester.ac.uk}

\vspace{10pt}
\begin{indented}
\item[]November 2018
\end{indented}

\begin{abstract}
The experimental thermonuclear reactor, ITER, is currently being constructed in
Cadarache, France. The reactor vessel will be constructred with a beryllium coated
wall, and a tungsten coated divertor. As a plasma-facing component, the divertor will
be under conditions of extreme temperature, resulting in the sputtering of tungsten
impurities into the main body plasma. Modelling and understanding the potential 
cooling effects of these impurities requires detailed collisional-radiative modelling.
These models require a wealth of atomic data for the various atomic species in the 
plasma. In particular, partial, final-state resolved dielectronic/radiative 
recombination (DR/RR) rate coefficients for tungsten are required. In this manuscript, we 
present our calculations of detailed DR/RR rate coefficients for the lanthanide-like,
and low ionization stages of tungsten, spanning charge states W$^{13+}$ to W$^{1+}$. The 
calculations presented here constitutes the first detailed exploration of such low
ionization state tungsten ions. We are able to reproduce the general trend of calculations
performed by other authors, but find significant differences between ours and their DR rate 
coefficients, especially at the lowest temperatures considered.

\end{abstract}

%
%
%
%
\ioptwocol

\section{Introduction}
Magnetically confined nuclear fusion is currently seen as humanity's best hope for
realising the prospect of near-limitless, clean energy. As population growth around
the world booms and energy needs grow inexorably larger, nuclear technologies are the
only energy generating mechanisms that can realistically meet these demands. One step
in realising nuclear fusion as an energy source is the upcoming thermonuclear 
experimental reactor, ITER, currently being built in Cadarache, France. First plasma
for ITER is currently scheduled for December 2025, with the first tritium-deuterium
campaign scheduled for 2035\footnote{https://www.iter.org/proj/inafewlines}. ITER
will be the first thermonuclear reactor to produce more energy than it consumes, with
a projected output of $Q=10$. The divertor, positioned at the separatarix of the 
reactor, is tasked with the removal of waste products and impurities, and will be
composed of several tungsten tiles. This particular metal has been chosen due to its
low affinity for tritium absorption, ability to withstand large power loads, and high
melting point. High-$Z$ atoms and ions are very efficient radiators due to synchrotron 
radiation (scales as the residual charge $z^2$). As tungsten is sputtered into the 
main body plasma from the divertor, the power loss from this radiation can cause 
cooling and potentially quenching of the reaction. Understanding how tokamak plasmas 
behave during normal operation, and when seeded with impurities requires the use of 
collisional-radiative modelling codes. A necessary ingredient for these models is a 
complete set of dielectronic and radiative recombination (DR/RR) rate coefficients 
for all elements in the plasma. In addition, this data needs to be partial rather 
than total to understand the level populations in the plasma, and also have to be 
final-state resolved. While this is relatively simple for ions with few electrons, 
it is no mean feat in the case of tungsten with 74 electrons.

The calculation of tungsten DR rate coefficient data is now well established as a 
subject of high priority, and many researchers have risen to the challenge. The first
isonuclear sequence calculation for tungsten was performed by Post~\etal 
\cite{post1977a,post1995a}, who used an average atom method as implemented by the 
ADPAK codes to calculate recombination rate coefficients (DR+RR). This data was used 
by P\"{u}tterich~\etal \cite{putterich2008a} to model observed spectral emission from 
the tokamak plasma at the ASDEX upgrade. To improve agreement with observation, 
P\"{u}tterich~\etal empirically scaled the recombination rate coefficients of 
W$^{20+}$-W$^{55+}$. Foster \cite{foster2008a} also calculated isonuclear DR rate 
coefficients using the Burgess General Formula \cite{burgess1965a}, and the 
Burgess-Bethe General Program \cite{badnell2003a}. The RR rate coefficients were 
approximated by scaling hydrogenic values. In addition, recombination rate 
coefficients have also been calculated using the FLYCHK code \cite{chung2005a}. Like
ADPAK, FLYCHK also uses an average atom method. The rate coefficients are currently 
hosted on the International Atomic Energy Agency's website
\footnote{https://www-amdis.iaea.org/FLYCHK/}. Despite the number of datasets 
available, poor agreement exists between all three of them. Further clarification is
needed. In an attempt to resolve the disagreement, \textit{The Tungsten Project} was 
created to calculate a set of final-state resolution DR and RR rate coefficients for 
tungsten using the distorted wave code {\sc autostructure} 
\cite{badnell1986a,badnell1997a,badnell2011a}. To date, the project has produced DR 
and RR rate coefficients for W$^{74+}$ - W$^{28+}$ 
\cite{preval2016a,preval2017b,preval2018a}. All data from 
\textit{The Tungsten Project} is currently hosted on the Atomic Data Analysis 
Structure (ADAS) website\footnote{http://www.open.adas.ac.uk} in the standard adf09
(DR) and adf48 (RR) formats, the specifications for which can also be found on the
ADAS website.

As well as large scale isonuclear sequence work, there are multiple detailed
calculations available that consider individual ions, or subsets of the isonuclear 
sequence. Such calculations are typically focused around ionization states that have 
closed or near-closed outer electron shells, and are performed using a multitude of 
codes. Using the Cowan \cite{cowanbook1981} and HULLAC \cite{barshalom2001a} codes, 
Safronova~\etal has calculated level resolved DR rate coefficients for W$^{5+}$, 
W$^{6+}$, W$^{28+}$, W$^{38+}$, W$^{45+}$, W$^{46+}$, W$^{63+}$, and W$^{64+}$ 
\cite{usafronova2012a,usafronova2012b,usafronova2011a,usafronova2016a,
usafronova2015a,usafronova2012c,usafronova2009a,usafronova2009b}. The Cowan code and
HULLAC have also been used by Behar~\etal calculate data for W$^{45+}$, W$^{46+}$, 
W$^{56+}$, and W$^{64+}$ \cite{behar1997a,behar1999a,behar1999b}, and Peleg~\etal for
W$^{56+}$ \cite{peleg1998a}. The Flexible Atomic Code (FAC) \cite{gu2003a} has also 
been used to calculate DR rate coefficients for tungsten. Li~\etal used FAC to 
calculate data for W$^{29+}$, W$^{39+}$, W$^{27+}$, W$^{28+}$, and W$^{64+}$ 
\cite{li2012a,li2014a,li2016a}, while Meng~\etal and Wu~\etal used the same code to 
calculate data for W$^{47+}$ \cite{meng2009a} and W$^{37+}$-W$^{+46}$ \cite{wu2015a} 
respectively. In addition, Kwon~\etal has also used FAC to calculate data for 
W$^{44+}$-W$^{46+}$ Most recently, Kwon \cite{kwon2018a} used FAC to calculate DR 
rate coefficients for the lanthanide isoelectronic ions of tungsten, spanning 
W$^{5+}$-W$^{10+}$.

One of the biggest difficulties in calculating DR rate coefficients for tungsten is
for ions with a half open $4f$-shell due to their complicated level structures. Multiple
storage ring experiments have been done to measure the DR rate coefficients of such 
ions. In particular, storage ring experiments such as those described in 
\cite{schippers2011a,spruck2014a,badnell2016a} concerned the W$^{18+}$-W$^{20+}$
ionization states. In an attempt to model the experimental data, Badnell~\etal used
an upgraded version of {\sc autostructure}, and several physical approximations which
are detailed in \cite{badnell2012a}. Reasonable agreement was seen for W$^{18+}$, but 
the inferred plasma rate coefficients for W$^{20+}$ were larger than the 
{\sc autostructure} values by a factor 3. It was concluded that the discrepancy was 
due to an insufficient amount of mixing being included in the calculation for this ion. 
Until computing facilities are able to handle the immense calculations involved, the 
onus appears to be on statistical methods to generate the necessary rate coefficients. 
Extensive work on these methods has been done by Dzuba~\etal \cite{dzuba2012a,dzuba2013a}, 
Berengut~\etal \cite{berengut2015a}, Harabati~\etal \cite{harabati2017a}, and Demura~\etal 
\cite{demura2017a}. A review on statistical methods for half open $4f$-shells is given
in Krantz~\etal \cite{krantz2017a}.

In 2015 the International Atomic Energy Agency (IAEA) convened a specialist meeting
to assess the quality of the data described above. This was done in terms of the 
codes used to calculate the data, the methods, and the agreement with other available
literature. The results and recommendations of the meeting were published in a 
detailed report by Kwon~\etal \cite{kwon2017a}.

As introduced by Preval~\etal \cite{preval2016a} ionization states of tungsten
considered in this paper will not be referred to by their isoelectronic sequence 
name. Instead, we opt to refer to the various ionization states by the number of 
valence electrons a particular state has. For example, H-like (one electron) is referred 
to as 01-like, Pd-like (46 electrons) is 46-like, and Ta-like (73 electrons) is 73-like.

The lanthanide sequence, plus the transition metals leading up to tantalum-like, 
consitutes the end of the isonuclear sequence of tungsten. The structure of these 
ions are relatively simple compared to sequences such as the open $4f$-shell ions. 
However, the difficulty in modelling these low charge ions lies in the positioning of 
resonances, as well as calculating a reliable atomic structure. Low charge ions 
will likely be observed at the divertor and scrape-off layer within ITER. 
The plasmas created at ITER will span a wide range of temperatures, 
ranging from 1eV at the divertor and scrape-off layer, to 40keV in the core. This paper concerns the calculation 
of partial, final-state resolved DR and RR rate coefficients for the 
lanthanide, and low charge sector of the tungsten isonuclear sequence, spanning 
61-like to 73-like (W$^{13+}$ to W$^{1+}$). The paper is laid out as follows: In Section 
\ref{theorysec} we present the theory underpinning our calculations. We then discuss 
our method, including the configurations included. Next, we show our results, and compare our 
rate coefficients to currently available literature. We then calculate an updated 
steady state ionization fraction including the data calculated in this work, and the 
data calculated in \cite{preval2016a,preval2017b,preval2018a}. Finally, we conclude 
the paper, and discuss future work on the open $4f$-shell of tungsten.

\section{Theory}\label{theorysec}
The theoretical framework has been described in previous works from 
\textit{The Tungsten Project}, and at length by Badnell~\etal \cite{badnell2003a},
however, we provide a brief summary here. All data described in this paper was
calculated using the distorted wave, atomic collision package {\sc autostructure}. The code
can calculate multiple atomic quantities including, but not limited to; energy 
levels, radiative rates, photoionization cross sections, and collision strengths. 
{\sc autostructure} solves the one particle kappa-averaged Dirac equation with a 
Thomas-Fermi potential. Energies from the code can be calculated in multiple resolutions, 
namely level resolution (intermediate coupling, IC), term resolution, or configuration 
resolution (configuration average, CA). The code is well established, and has been 
benchmarked in several experimental settings. Most recently, it was used to compare 
storage ring measurements of 04-like Ar (Ar$^{14+}$) \cite{huang2018a}.

Consider a target ion $X_{\nu}^{+z+1}$ with a residual charge $z$ in some initial state 
$\nu$, recombining with an electron into an ion $X_{\nu}^{+z}$ with final state $f$. The partial 
DR rate coefficient $^{DR}\alpha_{f\nu}^{z+1}$ for electron temperature $T_{e}$ can be written as
\begin{eqnarray}
^{DR}\alpha_{f\nu}^{z+1}(T_e)&=&\left(\frac{4{\pi}a_{0}^{2}I_{H}}{k_{B}T_{e}}\right)^{\frac{3}{2}}
                                \sum_{j}\frac{\omega_{j}}{2\omega_{\nu}}\exp{\left[-\frac{E}{k_{B}T_{e}}\right]} \nonumber \\
                                &\times&\frac{\sum_{l}{A_{j\rightarrow{\nu},E\,l}^a}{A_{j\rightarrow{f}}^{r}}}
                                {\sum_{h}{A_{j\rightarrow{h}}^{r}} + \sum_{m,l}{A_{j\rightarrow{m},E\,l}^{a}}},
\label{DReq}
\end{eqnarray}
where the $A^{r}$ and $A^{a}$ are the radiative and autoionization (Auger) rates 
respectively, $\omega_{\nu}$ and $\omega_{j}$ are the statistical weights for the 
$N$- and $(N+1)$-electron-ions respectively, and $E$ is the continuum electrons energy
minus its rest energy. The sum over $l$ covers the DR rate coefficient for each orbital 
angular momentum quantum number. The total radiative and Auger widths are calculated 
via the sums over $h$ and $m$. $I_{H}$ is the ionization energy of the hydrogen atom, 
$k_{B}$ is the Boltzmann constant, and $(4\pi{a_{0}^{2}})^{3/2}=6.6011\times{10}^{-24}$cm$^{3}$.

Using detailed balance, RR can be written in terms of its time-reverse process, 
photoionization. Therefore, the partial RR rate coefficient for a given $T_e$, 
$^{RR}\alpha_{f\nu}^{z+1}(T_e)$, can be calculated as
\begin{eqnarray}
^{RR}\alpha_{f\nu}^{z+1}(T_e) &=& \frac{c\,\alpha^3}{\sqrt{\pi}}
                                  \frac{\omega_{f}}{2\omega_{\nu}}\left(I_H k_B T_e\right)^{-3/2} \nonumber \\
                                  &\times&\int^\infty_0 E^2_{\nu f}\, {^{PI}\sigma_{\nu f}^{z}(E)}
                                  \exp{\left[-\frac{E}{k_{B}T_{e}}\right]}dE\,,
\label{RReq}
\end{eqnarray}
where $E_{\nu{f}}$ is the photon energy, and $c{\alpha}^{3}/\sqrt{\pi}=6572.67$ cm s$^{-1}$

In the case of very high temperatures, relativistic corrections to the Maxwell-Boltzmann
distribution need to be made. Known as the Maxwell-J\"{u}ttner \cite{synge1957a} 
distribution these corrections manifest as a simple multiplicative factor, which can
be expressed as 
\begin{equation}
F_{\mathrm{r}}(\theta) = \sqrt{\frac{\pi\theta}{2}} 
\frac{1}{K_{2}(1/\theta){\rm e}^{1/\theta}},
\end{equation}
where $K_{2}$ is the modified Bessel function of the second kind, 
$\theta=\alpha^2 k_{B}T/2I_H$, $\alpha$ is the fine structure constant, and $k_{B}$
is the Boltzmann constant. For the ions considered in this work, the correction 
factor is very close to unity due to the low peak abundance temperatures, however,
we continue to apply the factor to the rate coefficients presented in this paper to
maintain consistency with our previous work.

\section{Calculations}

\subsection{DR}
In previous publications concerning \textit{The Tungsten Project} the concept of a 
core excitation was used to simplify and reduce the computational task. In the case 
of the ions considered in this paper, while it is possible to extract the 
contributions of individual core excitations after calculation, it no longer makes 
sense to split the initial calculation in this way. This is because orbitals of 
higher eccentricity encroach on lower eccentricity orbitals. This results in orbitals 
such as $5s$, $5p$, $6s$, and others having lower energies than less eccentric orbitals 
such as $4f$, $5g$ etc. Therefore, as an electron radiates/autoionizes into the core, it 
will do so in a ``non-standard'' order.

In \cite{preval2016a,preval2017b,preval2018a} the configuration sets used to 
calculate DR included so called ``one-up, one-down'' configurations for mixing 
purposes. For the lanthanide series it is no longer necessary to include these 
configurations. This is because the mixing effects of single excitations are stronger 
than those from one-up, one-down configurations (see \cite{cowanbook1981}). We found 
that including these configurations  simply shifts the positions of the resonances at 
low temperature. Therefore, the benefit of keeping these configurations in the 
calculation was far outweighed by the benefit of computational effort saved. That 
said, three additional configurations are included both for mixing purposes, and to 
account for the $6s$ orbital being lower in energy than $5d$, $5f$, and $5g$. These 
configurations took the form $4f^{x}5s^{y}5p^{z-2}6s^{2}$, $4f^{x}5s^{y}5p^{z}6s6p$, 
and $4f^{x}5s^{y}5p^{z}6p^{2}$.

In Table \ref{table:coreconfig} we list the $N$-electron configurations used in our 
DR rate coefficient calculations for each ion, and the maximum principal quantum 
number $n$ and orbital angular momentum quantum number $\ell$ considered. We also 
indicate which core excitations we consider in this work. The $(N+1)$-electron 
configuration set is formed by simply adding an additional electron to the entire 
$N$-electron configuration set. For each charge state, DR from capture to 
Rydberg $n\ell$ states were calculated sequentially up to $n=25$, and then 
quasi-logarithmically up to $n=999$. The DR contributions from intermediate 
$n$-values were obtained using interpolation. For these calculations, we calculated 
DR for all $\ell$ values from $\ell=0$ to $\ell=18$. This is sufficient to 
numerically converge the total DR rate coefficient to $<1$\% over the entire ADAS 
temperature range, defined as $z^2(10-10^{7})$K, where $z$ is the charge residual of 
the ionization state being considered.

\subsection{RR}
RR provides the largest contribution to the total recombination rate coefficient for 
highly ionized species. As the residual charge of the ion decreases, RR gives way to
DR. Interestingly, RR makes a comeback in the low ionization states at lower
plasma temperatures. This is because in the case of low charged ions the high temperature
DR peaks are less separated in energy, meaning the smaller energy jumps are more important.
For RR, the rate coefficient scales more regularly with changes in residual charge.

The $N$-electron configuration set consisted of the ground state configuration of the 
ionization state being considered. The $(N+1)$-electron configuration set was formed 
by simply adding an additional electron to the $N$-electron configuration. 
Like DR, the contribution to RR from Rydberg $n\ell$ electrons was calculated 
sequentially for $n$ values up to $n=25$, and then quasi-logarithmically up to 
$n=999$. The contribution for intermediate $n$ was obtained using interpolation. We 
calculate contributions to RR explicitly for $\ell$ values from $\ell=0$ to 
$\ell=10$, and also include a non-relativistic top up to the RR-rate coefficient from 
$\ell=11$ to $\ell=150$ to numerically converge the RR rate coefficient to $<1$\% 
over the entire ADAS temperature range. 

\subsection{Optimisation}
Unlike multi-configuration Hartree-Fock codes, {\sc autostructure} optimises energy 
levels through the variation of scaling parameters $\lambda_{n\ell}$ as implemented 
in a Thomas Fermi potential $V_{TF}$. The parameters can be optimised so as to 
improve the agreement between theoretical and laboratory energy levels, or to
minimise the energy functional. There exist multiple algorithms with which to achieve 
these optimisations. However, these methods are far beyond the scope of this work, and
will not be discussed further.

In some cases, the ground state configurations of the ionization stage we considered 
did not agree with the accepted values listed on the NIST website. For our DR calculations,
when the ground state was correct, we left all $\lambda_{n\ell}$ set to 1.0. In the converse 
case, we varied a single $\lambda_{n\ell}$ as applied to all orbitals by hand until the ground 
state was in the correct position. This was done so as to maintain as close a consistency
with methods used in our previous works as possible. We summarise the $\lambda_{n\ell}$
values used in Table \ref{table:params}. For RR, we set $\lambda_{n\ell}$ to 1.0 for all
ionization states considered. A quick check showed that setting $\lambda_{n\ell}$ to the values
listed in Table \ref{table:params} did not affect the RR rate coefficients over the ADAS
temperature range. In addition, as we only include one N-electron configuration in the
calculation, the correct ground state as listed by NIST is found for all ionization states.

\section{Results}
\begin{figure}
\begin{centering}
\includegraphics[width=85mm]{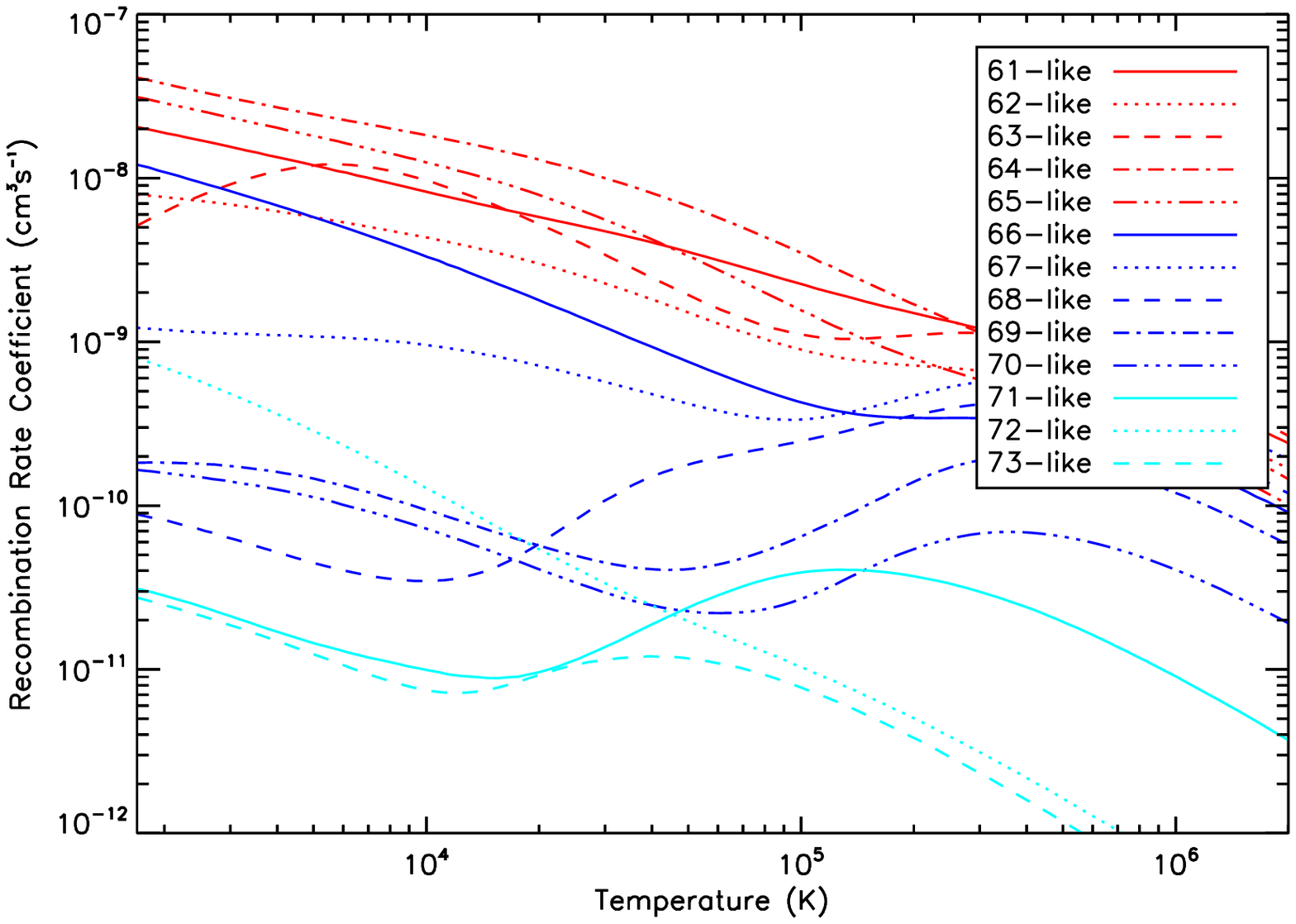}
\caption{Total DR rate coefficients calculated in level resolution (except for 63- 
and 71-like, calculated in configuration resolution) for 61- to 73-like.}
\label{fig:drtots}
\end{centering}
\end{figure}

\begin{figure}
\begin{centering}
\includegraphics[width=85mm]{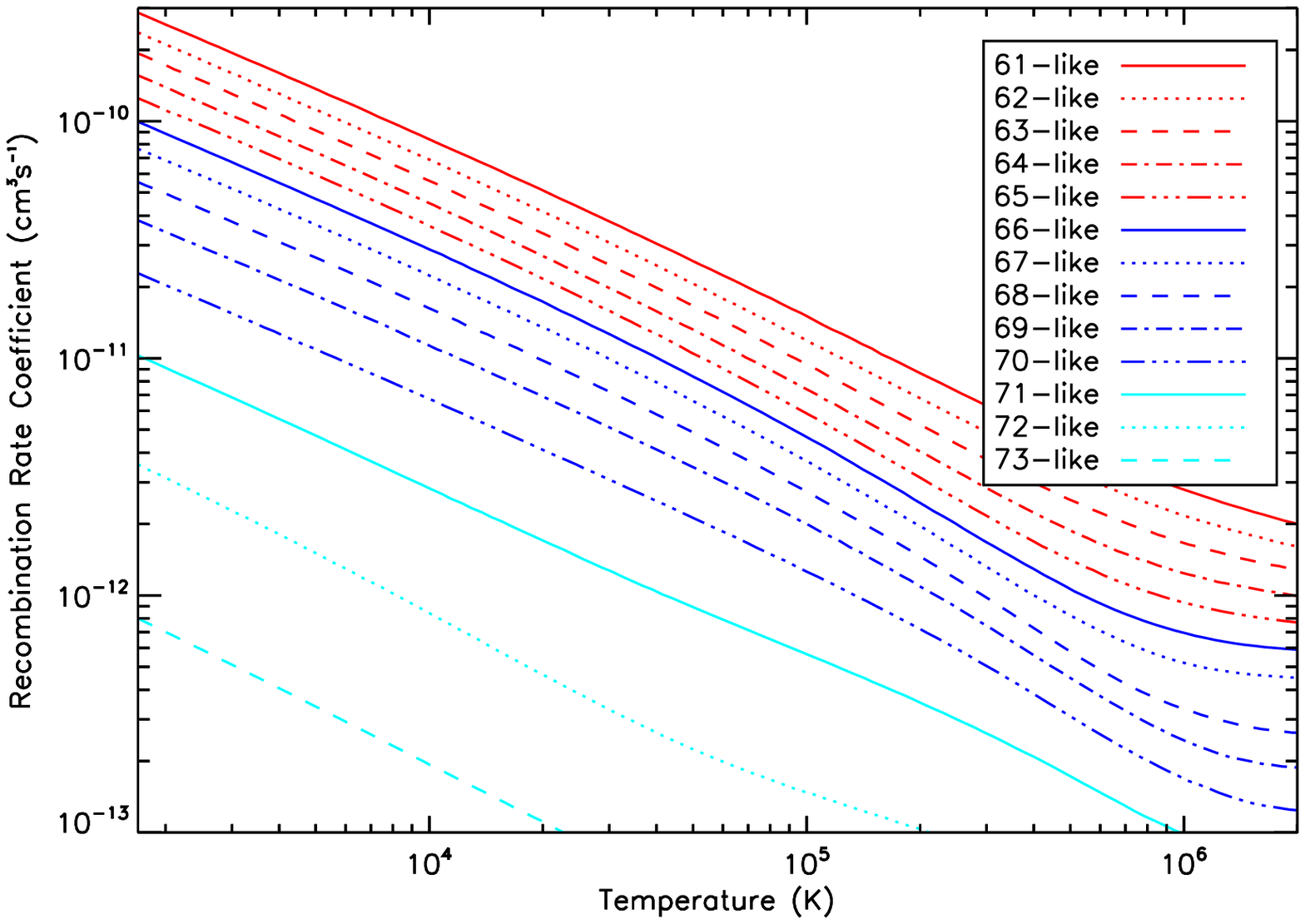}
\caption{Total RR rate coefficients calculated in level resolution (except for 63- 
and 71-like, calculated in configuration resolution) for 61- to 73-like.}
\label{fig:rrtots}
\end{centering}
\end{figure}

In this section we discuss the results of the calculations. In Figures \ref{fig:drtots} 
and \ref{fig:rrtots} we have plotted the total DR rate coefficients for 61- to 73-like 
in level resolution (except for 63- and 71-like, plotted in configuration resolution), 
and the total RR rate coefficients calculated in level resolution respectively. We 
consider each ionization state in turn, and compare the contribution of each core 
excitation to the total recombination rate coefficient. When comparing the core excitations, we also
indicate the peak fraction for that particular ion, calculated using the recombination rate
coefficients of P\"{u}tterich~\etal \cite{putterich2008a}, and the ionization rate coefficients 
of Loch~\etal \cite{loch2005a}. For each ionization state, we also indicate what value
of $\lambda_{n\ell}$ was used to produce the correct ground state as listed on the NIST website 
\cite{NIST_ASD2,kramida2006a,kramida2006b}. Note that odd parity states are indicated with an $*$ 
superscript in front of the level symbol.

\subsection{61-like: W$^{13+}$}
\begin{figure}
\begin{centering}
\includegraphics[width=85mm]{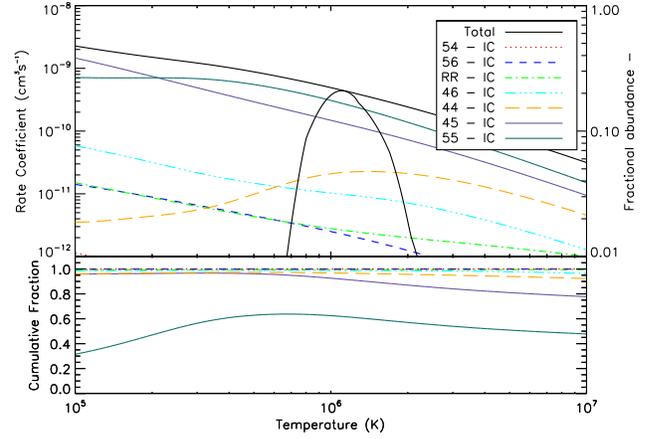}
\caption{Contributions to the total recombination rate coefficient for 61-like 
(top plot), and their cumulative fractions (bottom plot), calculated in IC. The method 
for calculating the cumulative fractions is given in text.}
\label{fig:61conts}
\end{centering}
\end{figure}

The ground state listed by NIST for this ion is $4f^{13}5s^{2}$ $^{2}F^{*}_{7/2}$. We set 
$\lambda_{n\ell}=0.99$ to reproduce this ground state in {\sc autostructure}. In Figure 
\ref{fig:61conts} we have plotted the individual contributions to the total 
recombination rate coefficients for 61-like, calculated in IC. The top plot shows the 
individual recombination rate coefficients as compared to the total, along with the 
peak abundance fraction (solid parabola) for 61-like. The bottom plot shows the 
cumulative contribution for each core excitation. This is calculated by summing the 
individual contributions, from largest to smallest, up to the core excitation being 
considered. This sum is then divided by the total recombination rate coefficient. The 
largest contribution to the recombination rate total comes from the 5--5 core 
excitation, contributing 62\% at peak abundance temperature ($1.16\times{10}^{6}$K).
This is followed by the 4--5 core excitation, which contributes 30\%. The remaining
10\% is comprised of 4--4, 4--6, RR, 5--6, and 5--4. The 5--5 core excitation is
strongest around the peak abundance. 5--5 decreases to 10\% of the recombination rate
total towards the lowest temperature considered (1690K). Towards higher temperatures,
5--5 decreases steadily, constituting 35\% of the total by $\sim{3}\times{10}^{8}$K.
The 4--5 core excitation is strongest at low temperatures, consituting 84\% of the 
recombination rate total at the lowest temperature considered. Towards higher 
temperatures, 4--5 decreases to 23\% of the total. RR only becomes significant 
towards higher temperatures, constituting 25\% of the recombination rate total at 
$\sim{3}\times{10}^{8}$K.

\subsection{62-like: W$^{12+}$}
\begin{figure}
\begin{centering}
\includegraphics[width=85mm]{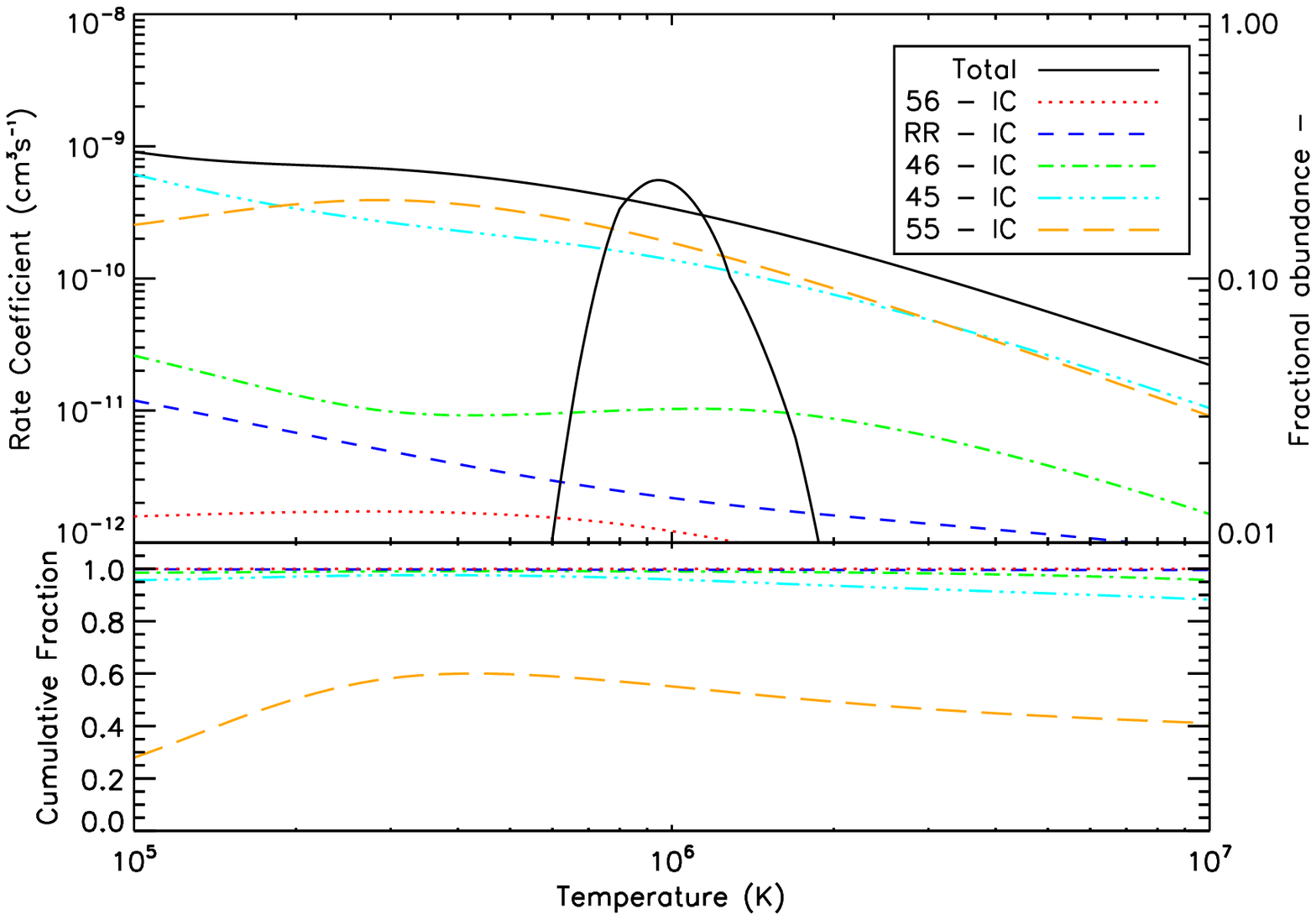}
\caption{Contributions to the total recombination rate coefficient for 62-like 
(top plot), and their cumulative fractions (bottom plot), calculated in IC.}
\label{fig:62conts}
\end{centering}
\end{figure}

The ground state given by NIST for this ion is $4f^{14}5s^{2}$ $^{1}S_{0}$. To reproduce
the correct ground state in {\sc autostructure} we set $\lambda_{nl}=0.98$. This ion was 
calculated in both IC and CA. In Figure \ref{fig:62conts} we have plotted the 
individual core excitations and their contributions to the recombination rate totals, 
along with their cumulative fractions. The largest contributions to the total comes 
from the 4--5 and 5--5 core excitations, contributing 41 and 55\% to the 
recombination rate total at peak abundance temperature ($9.4\times{10}^{5}$K)
respectively. This is followed by the 4--6 core excitation, which only contributes 
3\% to the total. The remainder of the total is composed of the 5--6 core excitation, 
and RR. Towards lower temperatures, the contribution of the 5--5 core excitation 
peaks at $\sim{4}\times{10}^{5}$K, consituting 60\% of the recombination rate total. 
The contribution from 5--5 then steadily decreases to 11\% at $\sim{20000}$K, 
followed by a slight increase to 23\% at $\sim{1000}$K. The contribution from 5--5 
decreases steadily with increasing temperature, constituting $\sim{30}$\% of the 
total by $\sim{1}\times{10}^{8}$K. The contribution from the 4--5 core excitation 
peaks at $\sim{17000}$K, consituting 84\% of the recombination rate total. Towards 
lower temperatures, the contribution from 4--5 decreases steadily from its peak 
value, constituting 73\% of the total by $\sim{1000}$K. Towards higher temperatures, 
the contribution from 4--5 again decreases steadily from it's peak value, 
constituting $\sim{40}$\% of the total by $\sim{1}\times{10}^{8}$K. The contribution 
from RR is very small below peak abundance temperatures, constituting only 3\% of the 
total at $\sim{1000}$K. However, there is a larger contribution from RR at the 
highest temperatures considered, constituting $\sim{20}$\% of the total at 
$\sim{1}\times{10}^{8}$K.

\subsection{63-like: W$^{11+}$}
\begin{figure}
\begin{centering}
\includegraphics[width=85mm]{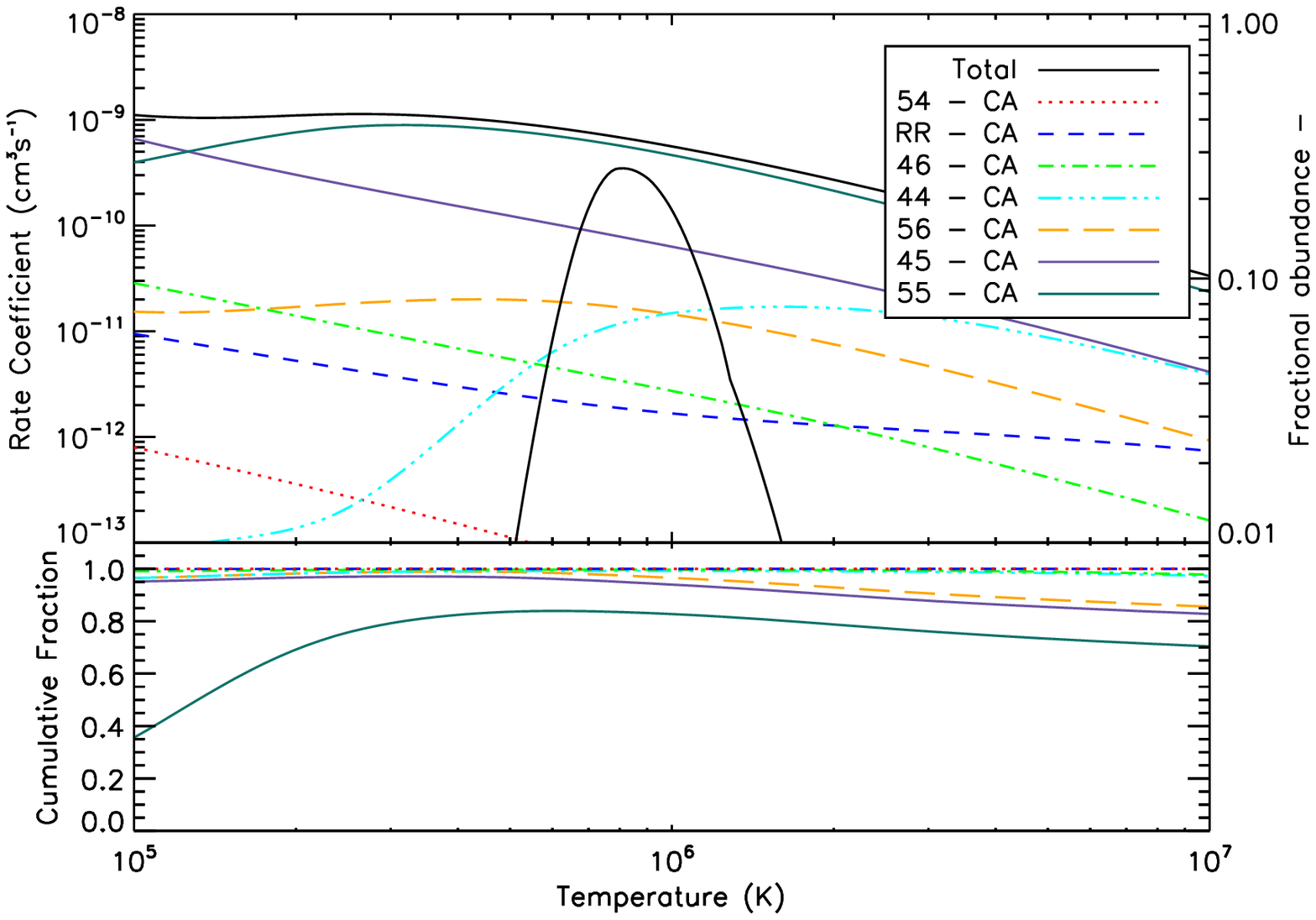}
\caption{Contributions to the total recombination rate coefficient for 63-like 
(top plot), and their cumulative fractions (bottom plot), calculated in CA.}
\label{fig:63conts}
\end{centering}
\end{figure}

This ionization state was calculated in CA only, as a representative calculation done
in IC is too computationally expensive at present. By setting $\lambda_{nl}=1.00$,
{\sc autostructure} predicts the ground state configuration to be $4f^{13}5s^{2}5p^{2}$ 
$^{4}F^{*}_{7/2}$. As this is in agreement with the value given by NIST, we do not alter $\lambda_{nl}$.
The individual contributions to the total recombination rate coefficient for this 
ionization state are plotted in Figure \ref{fig:63conts}, along with their cumulative 
fractions. The largest contributions to the total comes from the 4--5 and 5--5 
core-excitations, contributing $\sim{11}$ and $\sim{84}$\% to the total at peak 
abundance temperature ($8.1\times{10}^{5}$K) Towards lower temperatures, the 5--5 core
excitation gives way to 4-5, with the latter constituting $\sim{92}$\% of the recombination
rate total. 4--4 is the next largest contribution after 4--5 and 5--5, constituting
$\sim{13}$\% of the total at $\sim{3}\times{10}^{7}$K. The 4--6 core-excitation 
contributes very little for all temperatures, consituting a maximum of $\sim{3}$\% of the total at 
$\sim{7}\times{10}^{4}$K. The 5--4 core-excitation contributes $<1$\% for all temperatures, and 5--6 
only constitutes $\sim{3}$\% at maximum for temperatures $>1\times{10}^{6}$K. RR 
contributes little to the recombination rate total over all but the highest temperatures. 
At peak abundance temperature, RR contributes $<1$\%, whereas at the highest temperature 
considered ($2.4\times{10}^{8}$K), RR contributes $\sim{17}$\%. At low temperatures of 
$\sim{1000}$K, RR constitutes $\sim{7}$\% of the total. 

\subsection{64-like: W$^{10+}$}
\begin{figure}
\begin{centering}
\includegraphics[width=85mm]{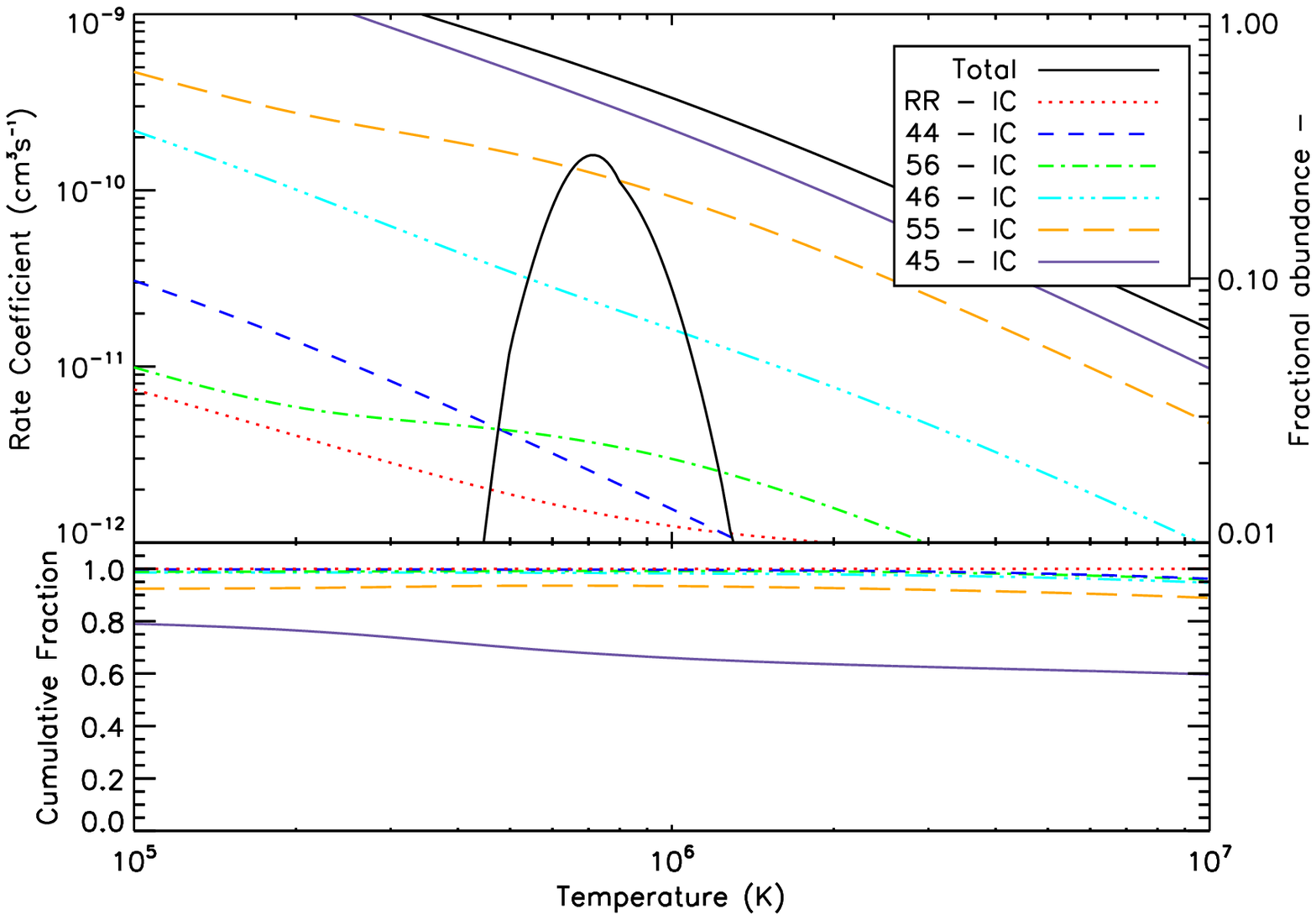}
\caption{Contributions to the total recombination rate coefficient for 64-like 
(top plot), and their cumulative fractions (bottom plot), calculated in IC.}
\label{fig:64conts}
\end{centering}
\end{figure}

The DR/RR rate coefficients for this ion were calculated in IC and CA. To reproduce
the ground state as given on the NIST website ($4f^{14}5s^{2}5p^{2}$ $^{3}P^*$), we 
set $\lambda_{n\ell}=0.99$. We
plot the individual contributions to the recombination rate total for 64-like in 
Figure \ref{fig:64conts}, along with their cumulative fractions. At peak abundance
temperature ($7.14\times{10}^{5}$K) the total is dominated by the contributions of the
4--5 and 5--5 core excitations, constituting 68 and 26\% to the total respectively. 
This is followed by the 4--6 core excitation, contributing 5\% to the total. 4--4, 
5--6, and RR contribute very little to the total at peak abundance temperature.
Combined, 4--4, 5--6, and RR constitute $<1$\% of the recombination rate total.
The contribution of the 4--5 core excitation decreases steadily with increasing 
temperature, constituting $\sim{50}$\% of the total by $\sim{1}\times{10}^{8}$K. The
converse is true for decreasing temperature, with the contribution from 4--5 constituting
90\% at the lowest ADAS temperature considered for this ion (1000K). The contribution
from 5--5 peaks at $\sim{4}\times{10}^{6}$K, constituting $\sim{30}$\% of the total.
For higher temperatures than this peak, the contribution from 5--5 decreases steadily, 
constituting 24\% of the total by $\sim{1}\times{10}^{8}$K. For lower temperatures
than the peak, the contribution from 5--5 again decreases steadily, constituting just
$\sim{2}$\% at 1000K. The contribution from 4--6 over the entire temperature range is
small, but non-negligible, constituting 4--7\% over the entire ADAS temperature range. 
The contribution from 4--4 and 5--6 is negligible at all temperatures. RR only 
becomes significant at the highest temperatures, constituting 20\% of the total at 
$\sim{1}\times{10}^{8}$K.

\subsection{65-like: W$^{ 9+}$}
\begin{figure}
\begin{centering}
\includegraphics[width=85mm]{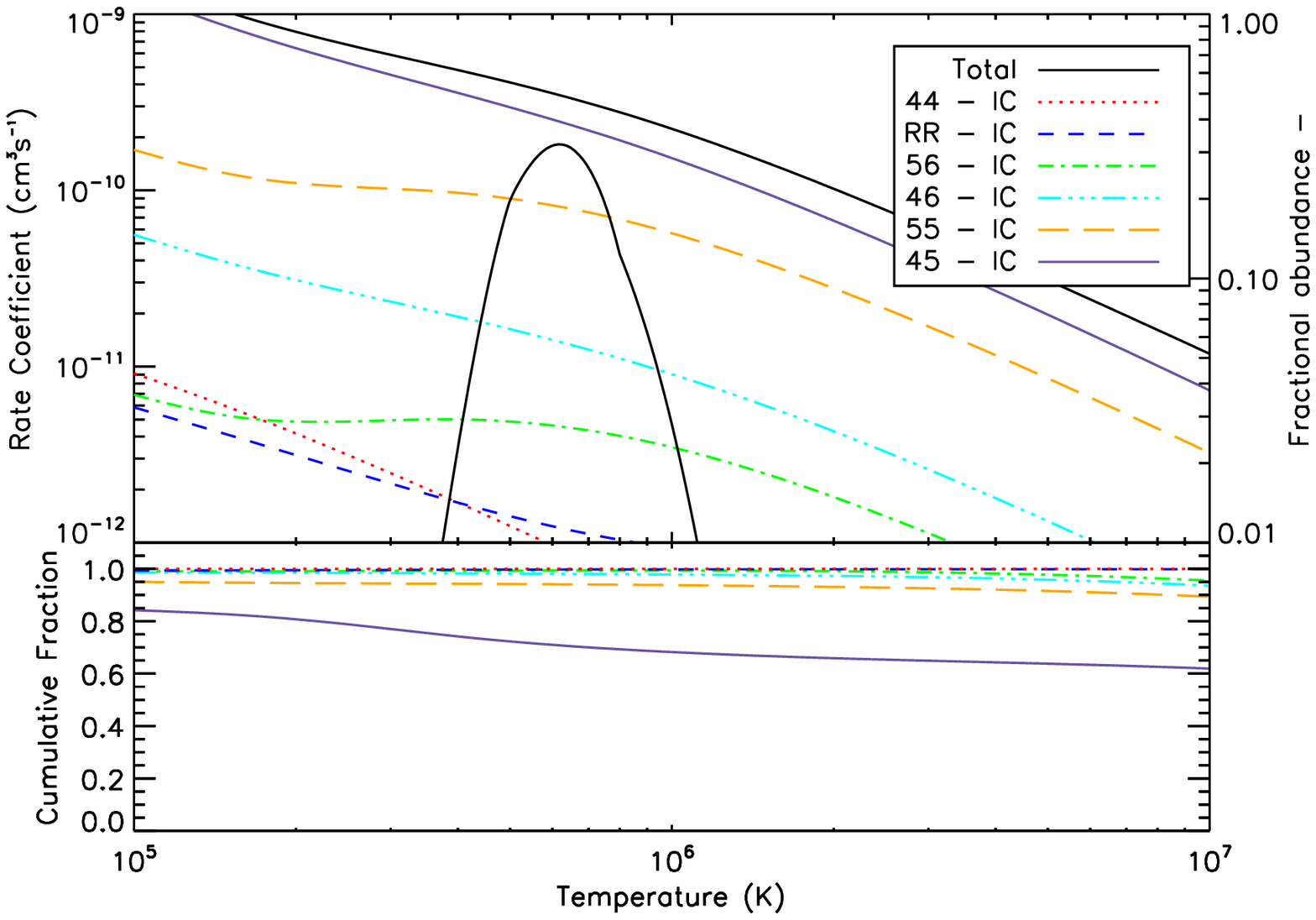}
\caption{Contributions to the total recombination rate coefficient for 65-like 
(top plot), and their cumulative fractions (bottom plot), calculated in IC.}
\label{fig:65conts}
\end{centering}
\end{figure}

For this ionization state, NIST lists the ground state configuration as being 
$4f^{14}5s^{2}5p^{3}$ $^{2}P^{*}_{3/2}$. To 
reproduce this ground state, we set $\lambda_{n\ell}=0.99$. In Figure 
\ref{fig:65conts} we have plotted the individual contributions to the recombination
rate total for 65-like, along with their cumulative fractions. The recombination rate
total is dominated by the 4--5 and 5--5 core excitations at peak abundance 
temperature ($6.18\times{10}^{5}$K), constituting 70 and 23\% of the total 
respectively. The contribution from 4--6 is smaller than in the case of 64-like,
constituting $\sim{4}$\% of the total. The 4--5 contribution 
decreases steadily with increasing temperature, constituting 50\% of the total at
$\sim{1}\times{10}^{8}$K. As with 64-like, the 4--5 contribution increases with
decreasing temperature, constituting 96\% of the total for the lowest ADAS temperature
considered for this ion (810K). The contribution from 5--5 peaks at $\sim{5}\times{10}^{6}$K,
constituting 28\% of the total. For increasing temperatures above this peak, 5--5 
decreases steadily, constituting 22\% at $\sim{1}\times{10}^{8}$K. Below this peak 
temperature, 5--5 decreases steadily with decreasing temperature, constituting just
3\% at the lowest ADAS temperature considered. The 4--6 contribution to the total
peaks at $\sim{5}\times{10}^{6}$K, constituting 4\% of the total. With increasing
temperature from this peak, the contribution from 4--6 decreases to 3\% by 
$\sim{1}\times{10}^{8}$K. The contribution from 4--6 becomes negligible with 
decreasing temperature below the peak. At maximum, the 5--6 core excitation contributes
2\% to the total. The 4--4 core excitation is completely negligible, and contributes 
$<1$\% to the total for the entire ADAS temperature range. RR again is only 
significant at high temperatures, constituting 22\% of the total at 
$\sim{1}\times{10}^{8}$K.

\subsection{66-like: W$^{ 8+}$}
\begin{figure}
\begin{centering}
\includegraphics[width=85mm]{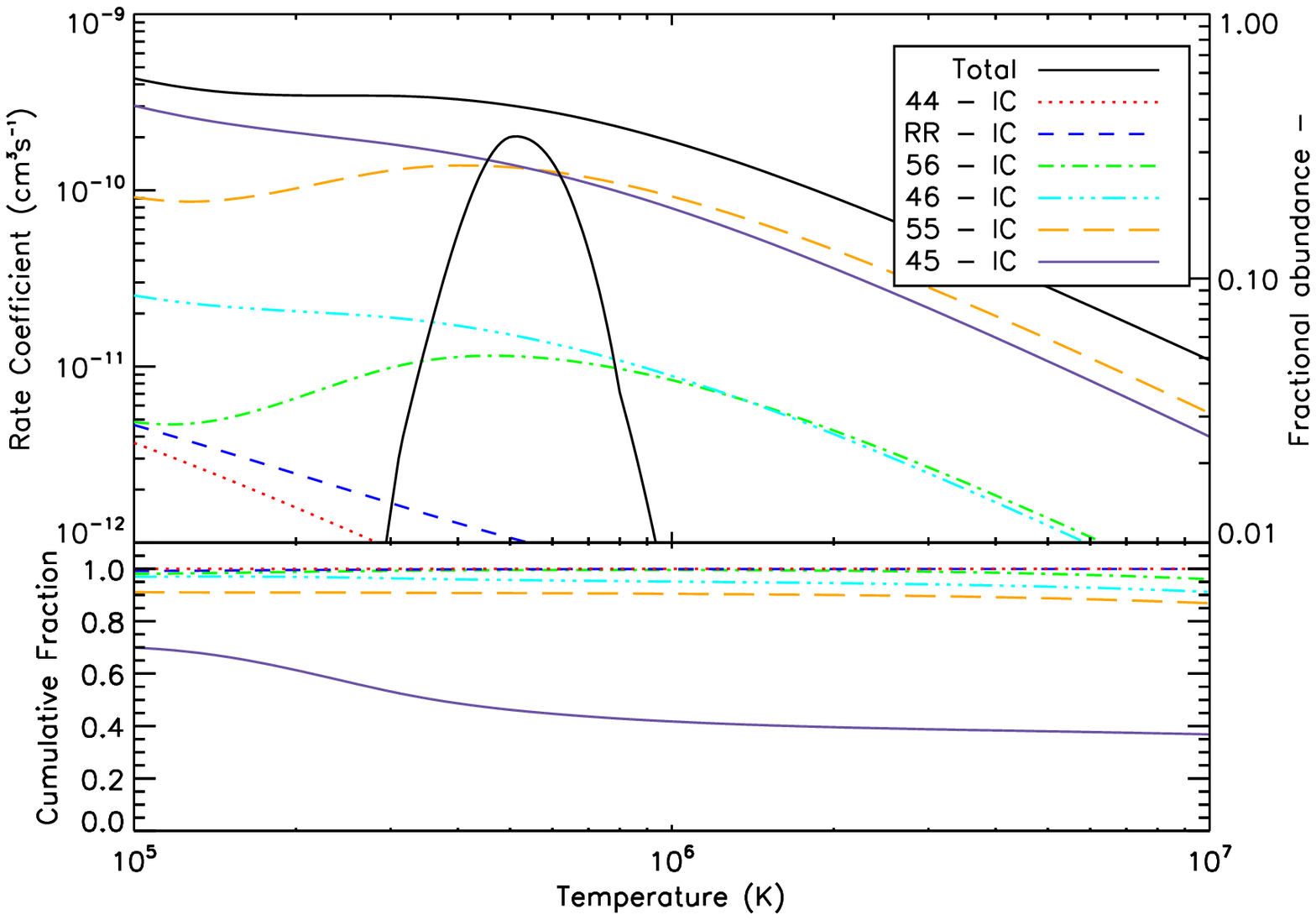}
\caption{Contributions to the total recombination rate coefficient for 66-like 
(top plot), and their cumulative fractions (bottom plot), calculated in IC.}
\label{fig:66conts}
\end{centering}
\end{figure}

The ground state for this ion is $4f^{14}5s^{2}5p^{4}$ $^{3}P_{2}$ In order to 
reproduce this in {\sc autostructure}, we set $\lambda_{n\ell}=0.99$. 
In Figure \ref{fig:66conts} we plot the
individual contributions to the recombination rate total, and their cumulative
fractions. At peak abundance temperature ($5.14\times{10}^{5}$K) the largest 
contributions to the total come from the 4--5 and 5--5 core excitations, constituting
46 and 45\% of the total respectively. These are followed by the 4--6 and 5--6 core
excitations, contributing 5 and 4\% to the total respectively. The 4--5 contribution
decreases with increasing temperature, constituting 30\% of the total at 
$\sim{1}\times{10}^{8}$K. With decreasing temperature, the contribution of 4--5 
increases steadily, peaking at $\sim{5000}$K, and constituting 71\% of the total.
The contribution from 4--5 then decreases slightly to 69\% of the total at the lowest
ADAS temperature considered (640K). The 5--5 contribution peaks at 
$\sim{4}\times{10}^{6}$K, constituting $\sim{50}$\% of the total. The contribution
from 5--5 then decreases steadily with increasing temperature above the peak, 
constituting $\sim{42}$\% of the total. With decreasing temperature below the peak,
the contribution from 5--5 again decreases, constituting 25\% of the total at 
$\sim{1000}$K. The contribution from 4--6 peaks at $\sim{2}\times{10}^{5}$K, 
constituting 6\% of the total. The contribution from 4--6 decreases with increasing
temperature above the peak, constituting 4\% at $\sim{1}\times{10}^{8}$K. With 
decreasing temperature below the peak, the contribution from 4--6 decreases steadily
to 5\% at $\sim{1000}$K. The 4--4 core excitation, as in the case of 65-like, is 
negligible, contributing $<1$\% for the entire ADAS temperature range. RR, again, is
only significant at high temperatures, constituting 20\% of the total at 
$\sim{1}\times{10}^{8}$K.

\subsection{67-like: W$^{ 7+}$}
\begin{figure}
\begin{centering}
\includegraphics[width=85mm]{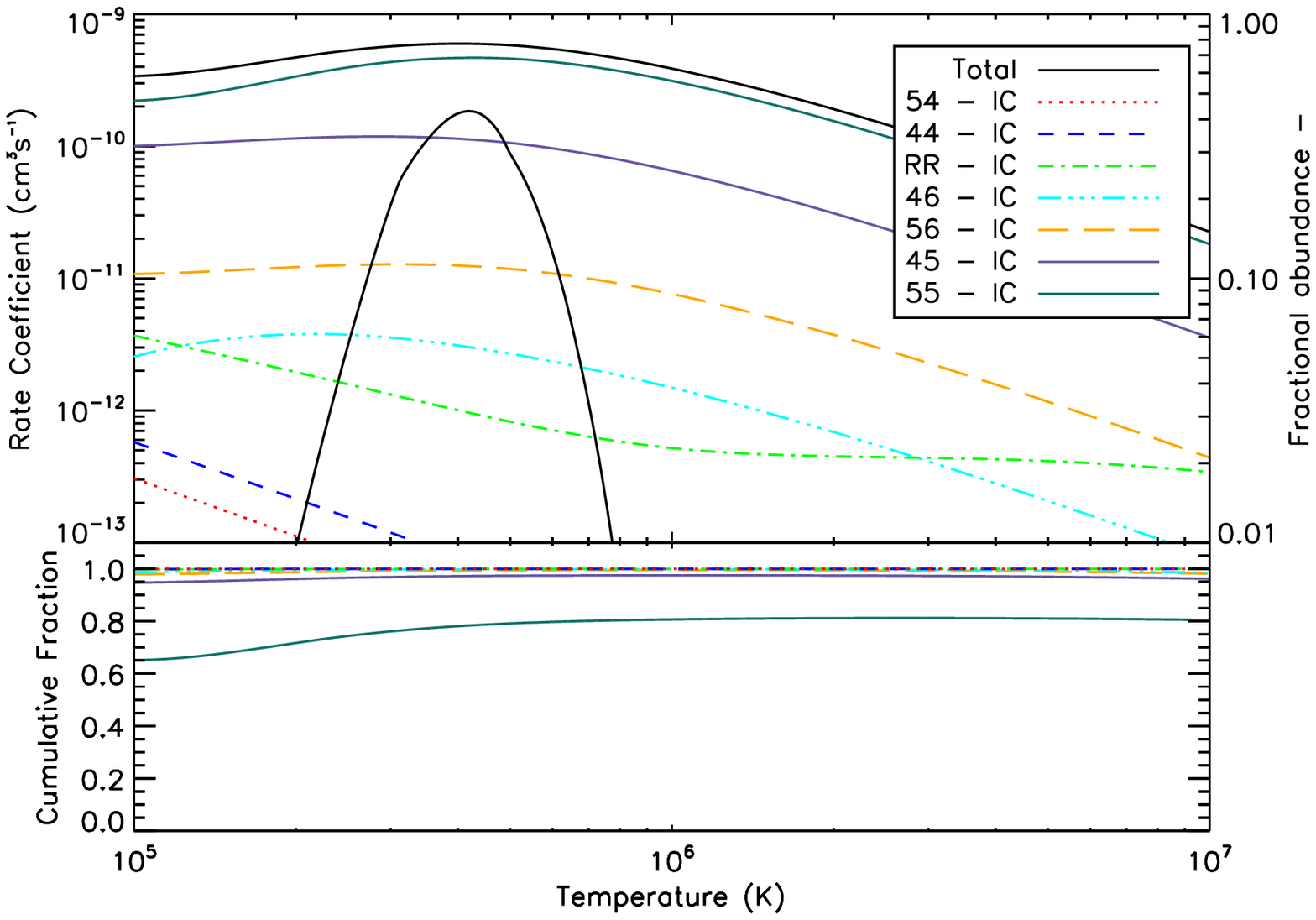}
\caption{Contributions to the total recombination rate coefficient for 67-like 
(top plot), and their cumulative fractions (bottom plot), calculated in IC.}
\label{fig:67conts}
\end{centering}
\end{figure}

For this ion, we set $\lambda_{nl}=0.98$ to reproduce the correct ground state as
listed on the NIST website ($4f^{13}5s^{2}5p^{6}$ $^{2}F^{*}_{7/2}$). We have plotted 
the individual contributions to the
total recombination rate coefficients for 67-like, as well as their respective 
cumulative fractions, in Figure \ref{fig:67conts}. The 5--5 core excitation is the
largest contributor, constituting 78\% of the total at peak abundance temperature 
($4.2\times{10}^{5}$K). This is followed by 4--5, contributing 19\% to the total, 
and 5--6, contributing 2\%. The contributions from 5--5 and 4--5 do not vary much 
with respect to temperature, with the contribution from 5--5 ranging from 
$\sim{60}$--80\% of the total, and the contribution from 4--5 ranging from 15-30\% 
of the total. The 5--6 contribution is small over the entire ADAS temperature range. 
At maximum, 5--6 contributes 3\% to the total at $\sim{10,000}$K. The 4--4, 5--4, 
and 4--6 core excitations are completely negligible. Combined, these core excitations 
constitute $<1$\% of the recombination rate total for the entire ADAS temperature range. 
RR again is significant at higher temperatures albeit much less so than in the case of 
preceeding ions, constituting just 6\% of the total at $\sim{5}\times{10}^{7}$K.

\subsection{68-like: W$^{ 6+}$}
\begin{figure}
\begin{centering}
\includegraphics[width=85mm]{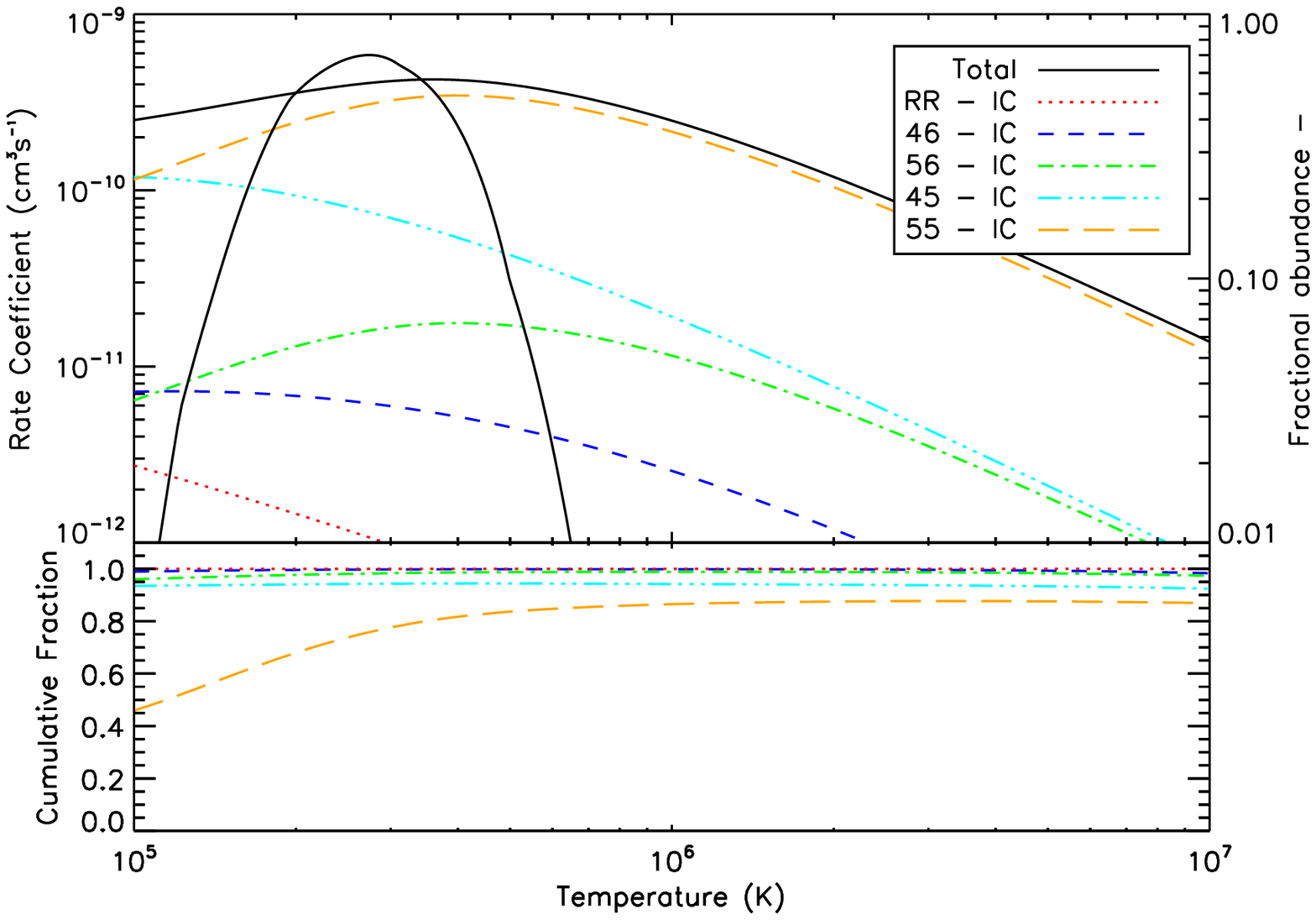}
\caption{Contributions to the total recombination rate coefficient for 68-like 
(top plot), and their cumulative fractions (bottom plot), calculated in IC.}
\label{fig:68conts}
\end{centering}
\end{figure}

We set $\lambda_{nl}=0.98$ for this ion to reproduce the correct ground state as
listed on the NIST website ($4f^{14}5s^{2}5p^{6}$ $^{1}S_{0}$). In Figure \ref{fig:68conts} 
we have plotted the individual contributions to the recombination rate total for 68-like, 
and their respective cumulative fractions. At peak abundance temperature 
($2.74\times{10}^{5}$K) the majority of the total recombination rate coefficient is composed of the 
5--5 and 4--5 core excitations, constituting 76 and 19\% of the total respectively. 
This is followed by 5--6 and 4--6, constituting 4 and 2\% of the total respectively.
The 5--5 core excitation contribution to the total has two maxima. The first is at
$\sim{1}\times{10}^{4}$K, constituting 62\% of the total, and the second is at 
$\sim{3}\times{10}^{6}$K, constituting 88\% of the total. Towards the lowest 
temperatures considered, the contribution from 5--5 deminishes, contributing 
$\sim{50}$\% to the total by $\sim{1000}$K. With increasing temperature, the 
contribution from 5--5 remains large, constituting 82\% of the total at 
$\sim{5}\times{10}^{7}$K. The 4--5 contribution peaks at $\sim{5}\times{10}^{6}$K, 
contributing 56\% to the total. With increasing temperature above this peak, 
the contribution from 4--5 decreases quite rapidly, constituting only 5\% of 
the total by $\sim{5}\times{10}^{7}$K. With decreasing temperature below the 
peak, the contribution from 4--5 again decreases rapidly, contributing 6\% at 
$\sim{1000}$K. The 5--6 core excitation does not contribute much over the ADAS 
temperature range, providing a peak contribution of 5\% at $\sim{5}\times{10}^{6}$K.
The contribution from 4--6 is the smallest for the entire ADAS temperature range,
contributing a maximum of $\sim{3}$\% to the total at $\sim{4}\times{10}^{4}$K.
As with the preceeding ions, the contribution from RR is negligible at peak abundance
temperature. However, RR becomes far more significant towards lower temperatures,
contributing 45\% to the total recombination rate coefficient at $\sim{1000}$K.

\subsection{69-like: W$^{ 5+}$}
\begin{figure}
\begin{centering}
\includegraphics[width=85mm]{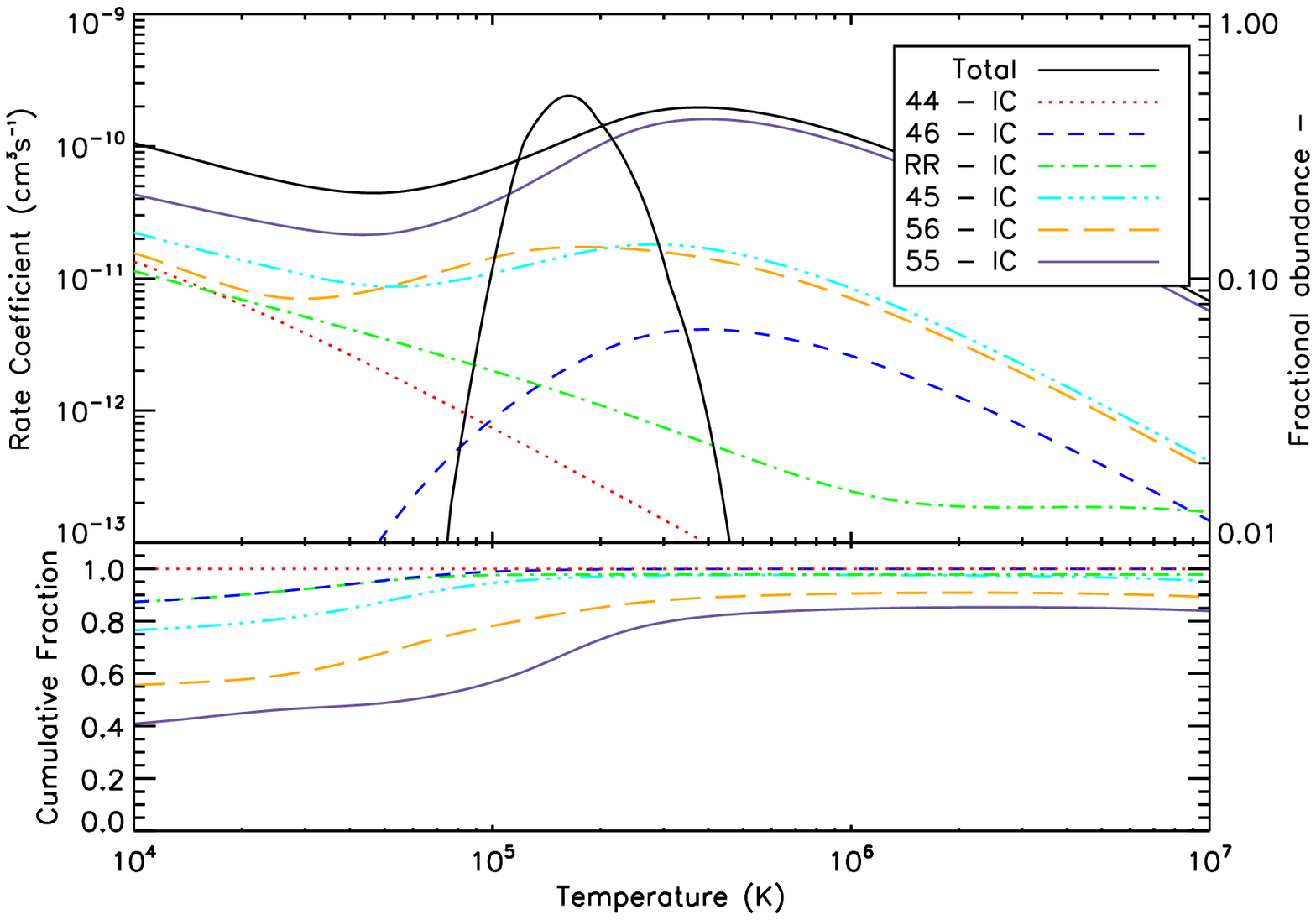}
\caption{Contributions to the total recombination rate coefficient for 69-like 
(top plot), and their cumulative fractions (bottom plot), calculated in IC.}
\label{fig:69conts}
\end{centering}
\end{figure}

For this ion, we set $\lambda_{nl}=0.97$ to reproduce the correct ground state
as listed on the NIST website ($5p^{6}5d$ $^{2}D_{3/2}$). In Figure \ref{fig:69conts} we 
have plotted the individual contributions to the total recombination rate coefficient for 69-like
calculated in IC, as well as their respective cumulative fractions. The 5--5 core 
excitation is dominant at peak abundance temperature ($1.63\times{10}^{5}$K), 
contributing 69\% to the recombination rate total. The next largest contribution 
comes from the 5--6 core excitation, contributing 15\% to the total. This is 
followed by 4--5, constituting 13\% of the total. The remainder of the total is
comprised of RR, 4--6, and 4--4, contributing a combined $\sim{3}$\%. 
The contribution from 5--5 peaks at $\sim{3}\times{10}^{6}$K, contributing
85\% to the total. For increasing temperature above this peak, the contribution
decreases slightly, constituting 77\% of the total at $\sim{5}\times{10}^{7}$K.
For decreasing temperature below the peak, the contribution from 5--5 decreases
gradually, constituting 25\% of the total by $\sim{1000}$K. For temperatures 
$\le{1}\times{10}^{5}$K the contribution from the 5--6 core excitation is 
relatively steady, constituting 10-20\% of the total. Above this temperature, the 
contribution from 5--6 decreases slowly, constituting 5\% of the total at 
$\sim{5}\times{10}^{7}$K. The contribution from the 4--5 core excitation peaks at 
$\sim{3000}$K, contributing 24\% to the total. For increasing temperature above 
this peak, the contribution from 4--5 decreases slowly, constituting 6\% at 
$\sim{5}\times{10}^{7}$K. For decreasing temperature below the peak, the contribution 
from 4--5 decreases rapidly, contributing $<1$\% at the lowest ADAS temperature considered 
(250K). The 4--6 core excitation contributes very little over the entire ADAS temperature 
range.At maximum, 4--6 contributes 2\% to the total, but only for temperatures 
$\ge{1}\times{10}^{5}$K The 4--4 core excitation is only significant for low temperatures. 
At $\sim{1000}$K 4--4 contributes 21\% to the total. The contribution from RR is more 
significant towards lower temperatures, contributing 24\% to the total at $\sim{1000}$K. 
However, at higher temperatures, RR is non-negligible, contributing 11\% to the total 
at $\sim{5}\times{10}^{7}$K.

\subsection{70-like: W$^{ 4+}$} 
\begin{figure}
\begin{centering}
\includegraphics[width=85mm]{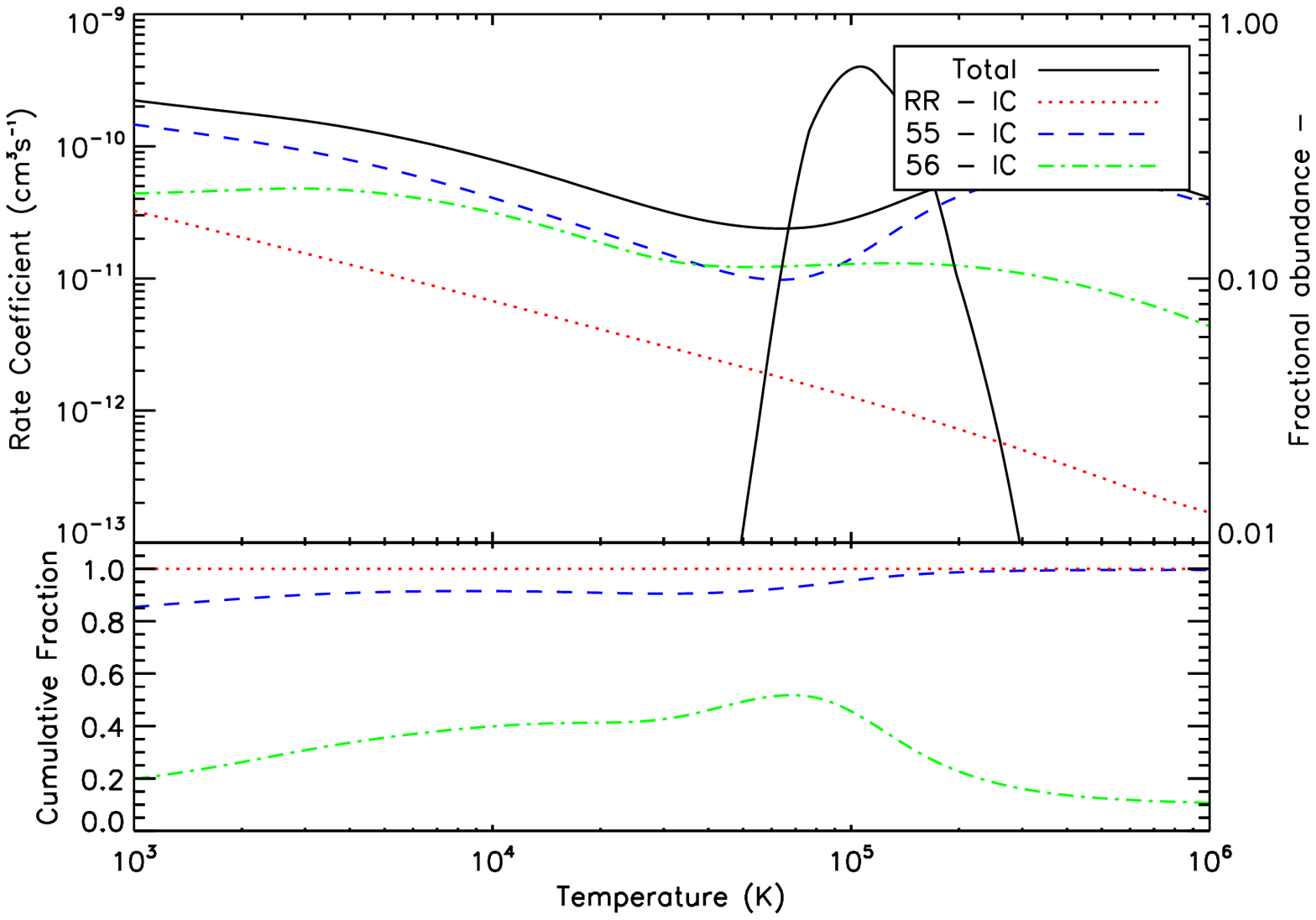}
\caption{Contributions to the total recombination rate coefficient for 70-like 
(top plot), and their cumulative fractions (bottom plot), calculated in IC.}
\label{fig:70conts}
\end{centering}
\end{figure}

To date, no detailed calculations of DR rate coefficients have been performed for this
ion up to 73-like. Thus, the present work constitutes the first such calculations. In Figure 
\ref{fig:70conts} we have plotted the recombination rate coefficients for 5-5, 5-6, and RR 
for 70-like, calculated in IC. Setting $\lambda_{nl}=0.96$ gives the correct ground 
state of $5p^{6}5d^{2}$ $^{3}F_{2}$ as listed on NIST. The contributions from 5-5 and 5-6 are 
comparable at peak abundance temperature ($7.0\times{10}^{4}$K), constituting 41 and 
51\% of the recombination rate coefficient total respectively. Interestingly, the 
peak abundance temperature marks the largest contribution to the recombination rate 
total for 5-6. For temperatures greater than the peak abundance temperature, the 
contribution from 5-6 decreases steadily. By $\sim{3}\times{10}^{7}$K, 5-6 contributes 
only 8\% to the recombination rate total. Likewise, for temperatures less than the peak 
abundance temperature, the contribution from 5-6 steadily decreases to 2\% at $\sim{200}$K. 
The largest contribution from 5-5 occurs at $\sim{10}^{6}$K, constituting $\sim{90}$\% 
of the recombination rate total. For temperatures $\ge{10}^{6}$K the contribution from
5-5 decreases steadily from its maximum to $\sim{76}$\% at $\sim{3}\times{10}^{7}$K.
For temperatures $\le{10}^{6}$K the 5-5 contribution decreases to 41\% at the peak
abundance temperature, after which the contribution begins increasing again, constituting
63\% of the total at $\sim{200}$K. RR contributes little at peak abundance temperature,
constituting 7\% of the recombination rate total. For temperatures greater than the peak abundance
temperature, the contribution from RR decreases slightly, and then gradually increases with 
increasing temperature, constituting 16\% of the recombination rate total at $\sim{3}\times{10}^{7}$K.
For temperatures less than the peak abundance temperature, the contribution from RR increases 
gradually, constituting 34\% of the total at the lowest temperature considered for this ion
($\sim{200}$K).

\subsection{71-like: W$^{ 3+}$}
\begin{figure}
\begin{centering}
\includegraphics[width=85mm]{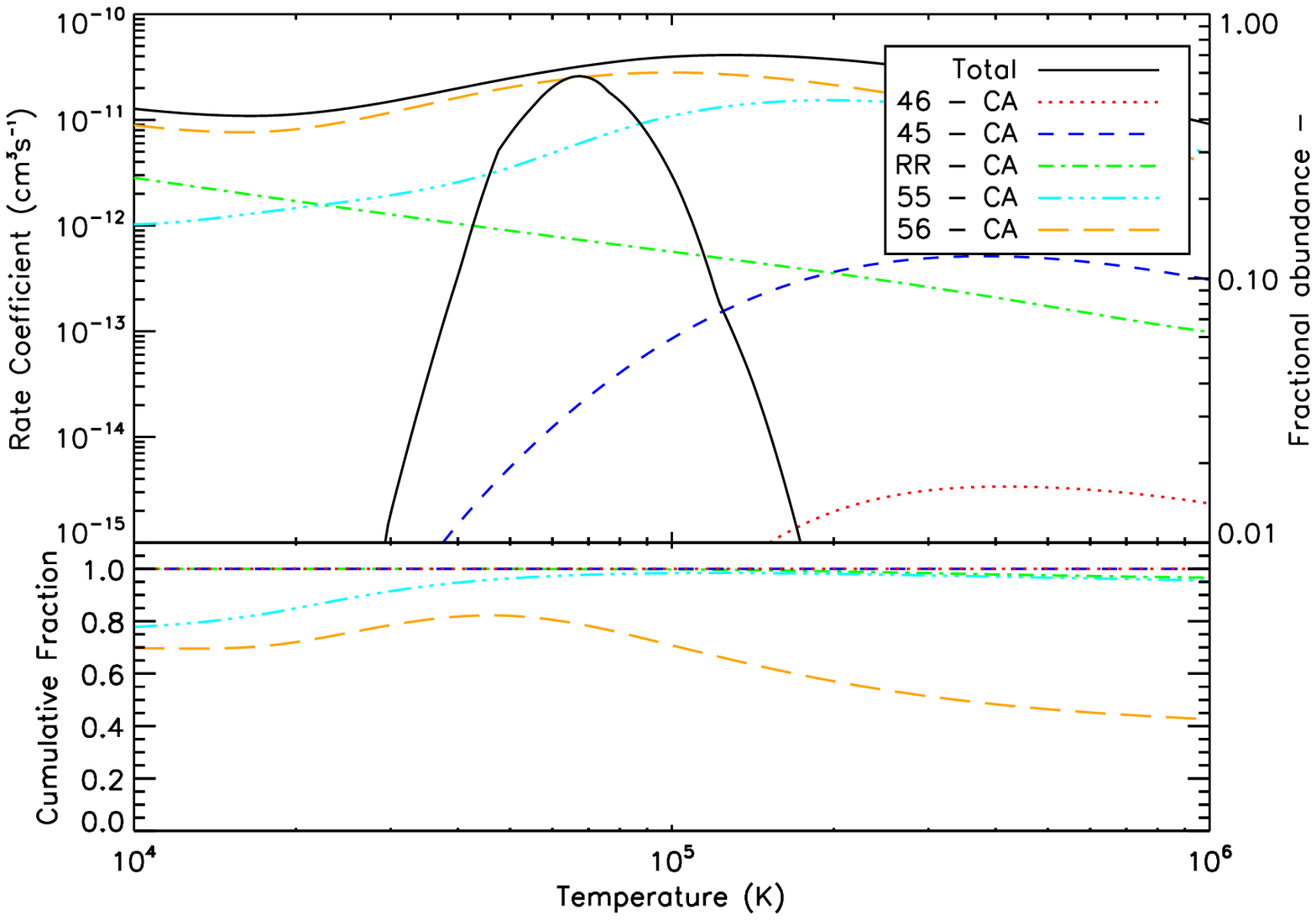}
\caption{Contributions to the total recombination rate coefficient for 71-like 
(top plot), and their cumulative fractions (bottom plot), calculated in CA.}
\label{fig:71conts}
\end{centering}
\end{figure}

The recombination rate coefficients for this ionization state were calculated in CA only,
and are plotted in FIgure \ref{fig:71conts}. We set $\lambda_{nl}=0.96$ to reproduce the 
correct ground state as listed on NIST ($5p^{6}5d^{3}$ $^{4}F_{3/2}$). The largest 
contributions to the recombination rate total come from the 5-5 and 5-6 core excitations, 
contributing $\sim{28}$ and 70\% to the total at peak temperature respectively. The 4-5 
and 4-6 core excitations contribute very little to the total over the entire ADAS temperature 
range. The 4-5 core excitation contributes $\sim{3-4}$\% to the total for temperatures 
$>7\times{10}^{5}$K, while the 4-6 core excitation contributes $<1$\% for all temperatures. 
At temperatures $>1.8\times{10}^{7}$K, 5-5, 5-6, and RR contribute to the total equally, composing
$\sim{33}$\% of the recombination rate total. Towards the lowest temperature considered for this
ion (90K), the recombination rate total is dominated by RR, constituting 100\% of the total.

\subsection{72-like: W$^{ 2+}$}
\begin{figure}
\begin{centering}
\includegraphics[width=85mm]{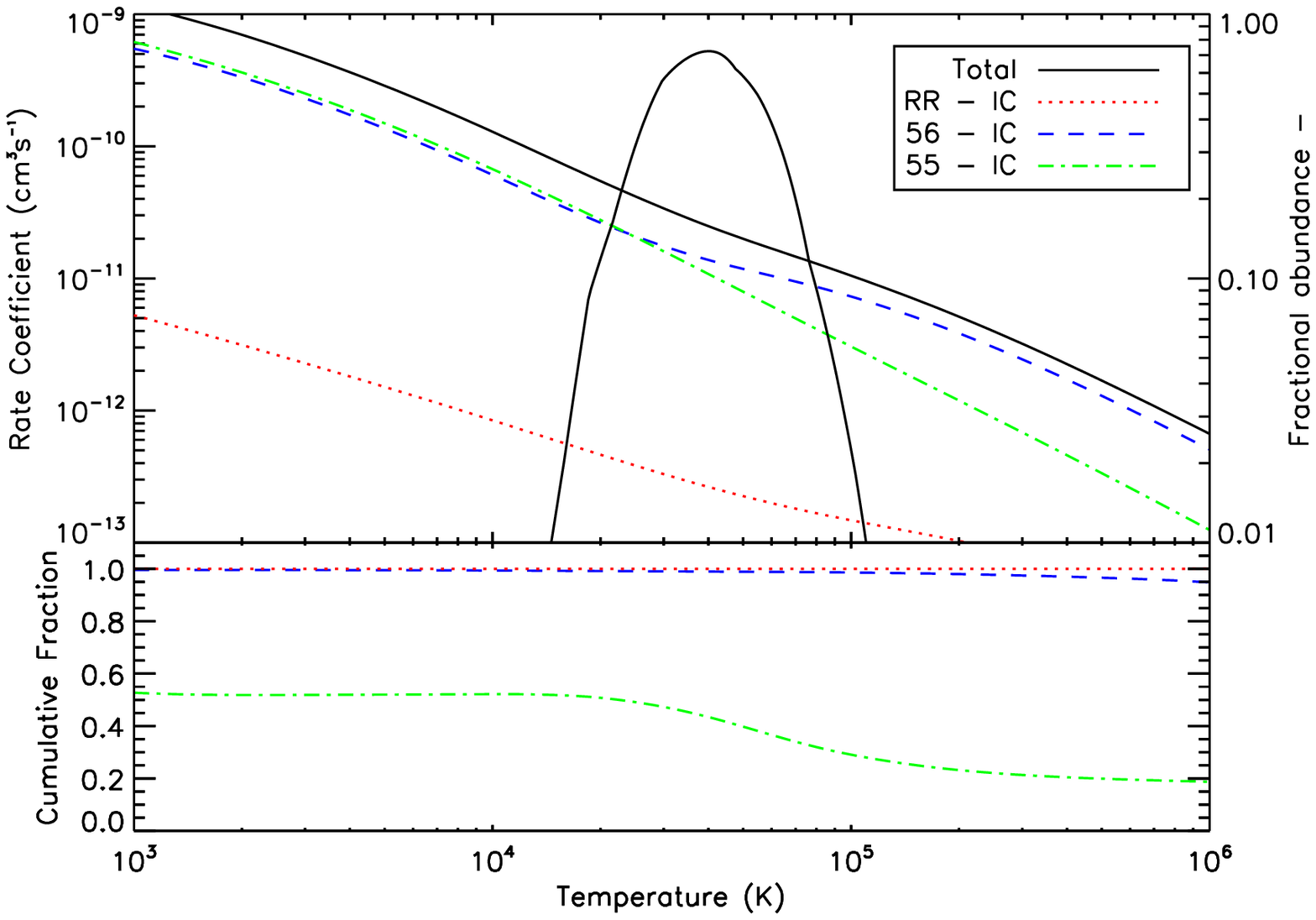}
\caption{Contributions to the total recombination rate coefficient for 72-like 
(top plot), and their cumulative fractions (bottom plot), calculated in IC.}
\label{fig:72conts}
\end{centering}
\end{figure}

In Figure \ref{fig:72conts} we have plotted the individual contributions to the total recombination rate 
coefficients for 72-like, along with their cumulative fractions. To reproduce the ground state listed on
NIST ($5p^{6} 5d^{4}$ $^{5}D_{0}$), we set $\lambda_{nl}=0.96$. This value also improves the 
general agreement between the {\sc autostructure} energies, and the values listed on NIST.
At peak abundance temperature ($2.9\times{10}^{4}$K), the contributions from 5--5 and 5--6 
are nearly equal, constituting 47 and 52\% of the recombination rate total respectively. The 5--5
core excitation provides the largest contribution at $\sim{200}$K, constituting 61\% of the recombination
rate total. For temperatures $<200$K, 5--5 decreases slightly, constituting 48\% of the recombination
rate total at $\sim{40}$K. Caution should be taken in interpreting this result, as low temperature DR
is very sensitive to the positioning of resonances at threshold. The 5--5 contribution steadily decreases 
for increasing temperature, and constitutes only $\sim{10}$\% of the recombination rate total at
$\sim{8}\times{10}^{6}$K. For 5--6, the largest contribution occurs at $\sim{5}\times{10}^{5}$K,
constituting 77\% of the recombination rate total. This decreases to 38\% at $\sim{200}$K, and then
increases slightly to 51\% at $\sim{40}$K. It is interesting to note that the contribution from 5--6 is
larger than that of 5--5 at low temperatures. As the $6s$ orbital encroaches upon the $5d$ orbital,
it is energetically more favourable for transitions to $6s$ to occur at low temperatures. 
The contribution from RR is small for temperatures $<1.8\times{10}^{6}$K, constituting $<10$\%
of the recombination rate total. For higher temperatures, the contribution from RR increases 
gradually to 47\%, becoming comparable to the 5--6 contribution.

\subsection{73-like: W$^{ 1+}$}
\begin{figure}
\begin{centering}
\includegraphics[width=85mm]{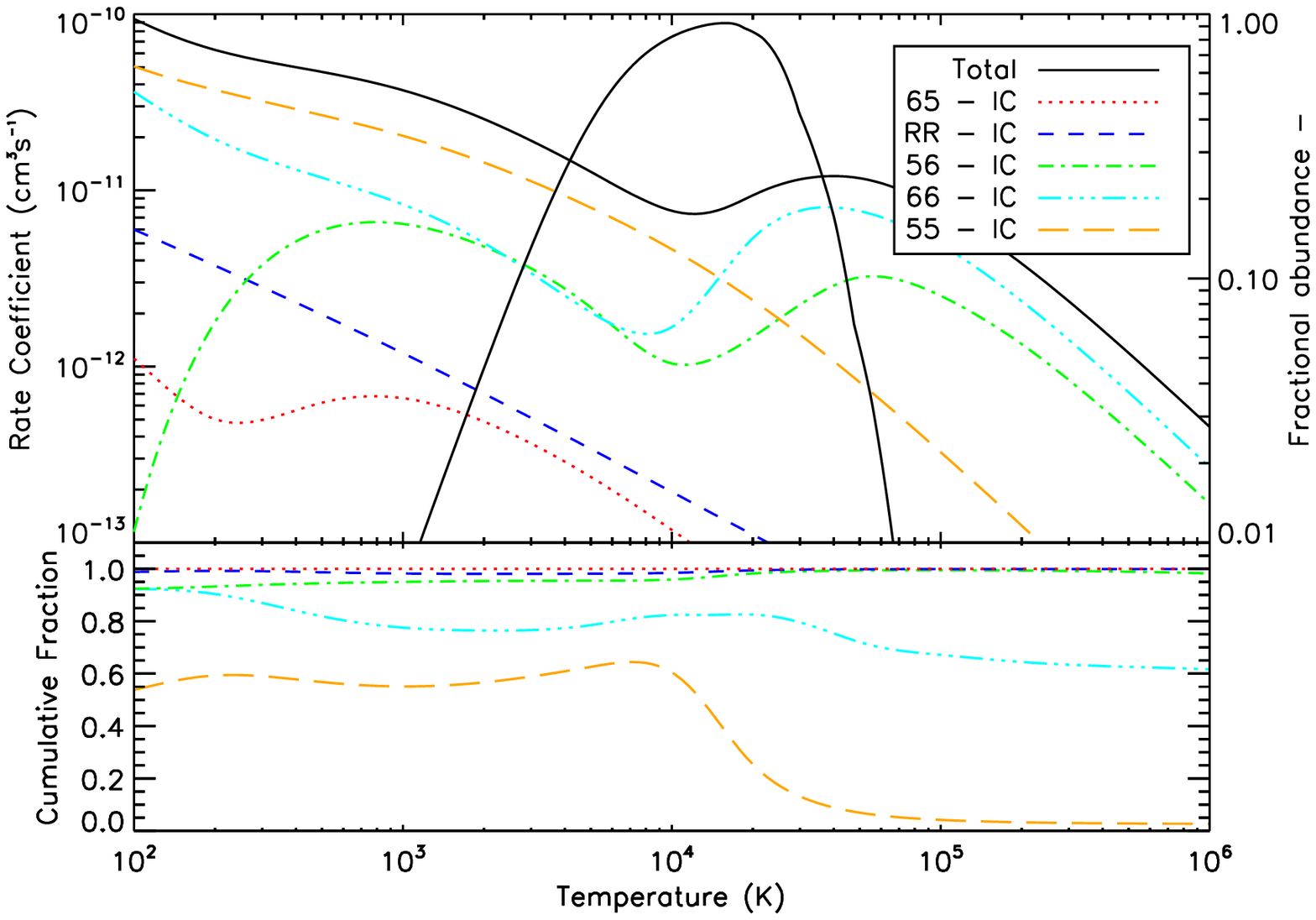}
\caption{Contributions to the total recombination rate coefficient for 73-like 
(top plot), and their cumulative fractions (bottom plot), calculated in IC.}
\label{fig:73conts}
\end{centering}
\end{figure}

We have plotted the individual contributions to the recombination rate total for 73-like in
Figure \ref{fig:73conts}, calculated in IC. We set $\lambda_{nl}=0.96$, to reproduce the correct
ground state as listed by NIST ($5d^{4}6s$ $^{6}D_{1/2}$). This also improves the general agreement 
between the {\sc autostructure} energies, and those listed on the NIST website. At peak abundance 
temperature ($1.6\times{10}^{4}$K) there is an interesting competition between the 5--5, 
5--6, and 6--6 core excitations. The 5--5 core excitation experiences a sharp drop in its 
contribution to the recombination total, constituting $\sim{60}$\% of the total at 
$\sim{10,000}$K, followed by a rapid drop to just $\sim{10}$\% at 37,000K. This is a 
result of the 5--6 and 6--6 DR rate coefficients peaking at 40,000 and 60,000K respectively, 
corresponding to the $5d\rightarrow{6s}$ and $6s\rightarrow{6p}$ promotions respectively. 
The sharp drop in the contribution is compensated for by 6--6, contributing $\sim{20}$\% at 
$\sim{10,000}$K, followed by a sharp rise to $\sim{65}$\% by $\sim{37,000}$K. The contribution 
from 5--6 is fairly constant, constituting 15\% of the total at peak abundance temperature. 
The contribution from 5--5 decreases steadily with increasing temperature, constituting just 
3\% at $\sim{1}\times{10}^{6}$K. Below peak abundance temperature, the contribution remains 
constant with decreasing temperature, constituting $\sim{50}$\% of the total. The 6--5 core 
excitation is negligible, and contributes $<2$\% over the entire ADAS temperature range.
The contribution from RR is small for all temperatures. At maximum, RR contributes 6\% to the
total at $\sim{100}$K.

\section{Comparison with other works}\label{authorcomp}
As discussed in the introduction, the lanthanide series and beyond are relatively
unexplored areas in terms of DR rate coefficient calculations. The only available
detailed calculations are from Safronova~\etal covering W$^{5+}$-W$^{6+}$ 
\cite{usafronova2012a,usafronova2012b}, and Kwon covering W$^{5+}$-W$^{10+}$ \cite{kwon2018a}. 
The DR rate coefficients from Kwon were calculated using FAC \cite{gu2003a}, while the DR rate 
coefficients from Safronova~\etal were calculated using a combination of {\sc hullac} \cite{barshalom2001a}, 
and the Cowan atomic structure code. In the absence of experimental data (such as those from
storage ring experiments), comparisons between DR rate coefficients calculated using different codes
offer an alternative method with which to benchmark these data.
In general, the DR rate coefficients for a particular ion at low temperatures can be 
highly uncertain due to the positioning of resonances. This is especially so in the 
case of low-ionization state tungsten, as calculating an accurate structure for such 
ions can be prohibitively difficult.

\subsection{64-like: W$^{10+}$}
\begin{figure}
\begin{centering}
\includegraphics[width=85mm]{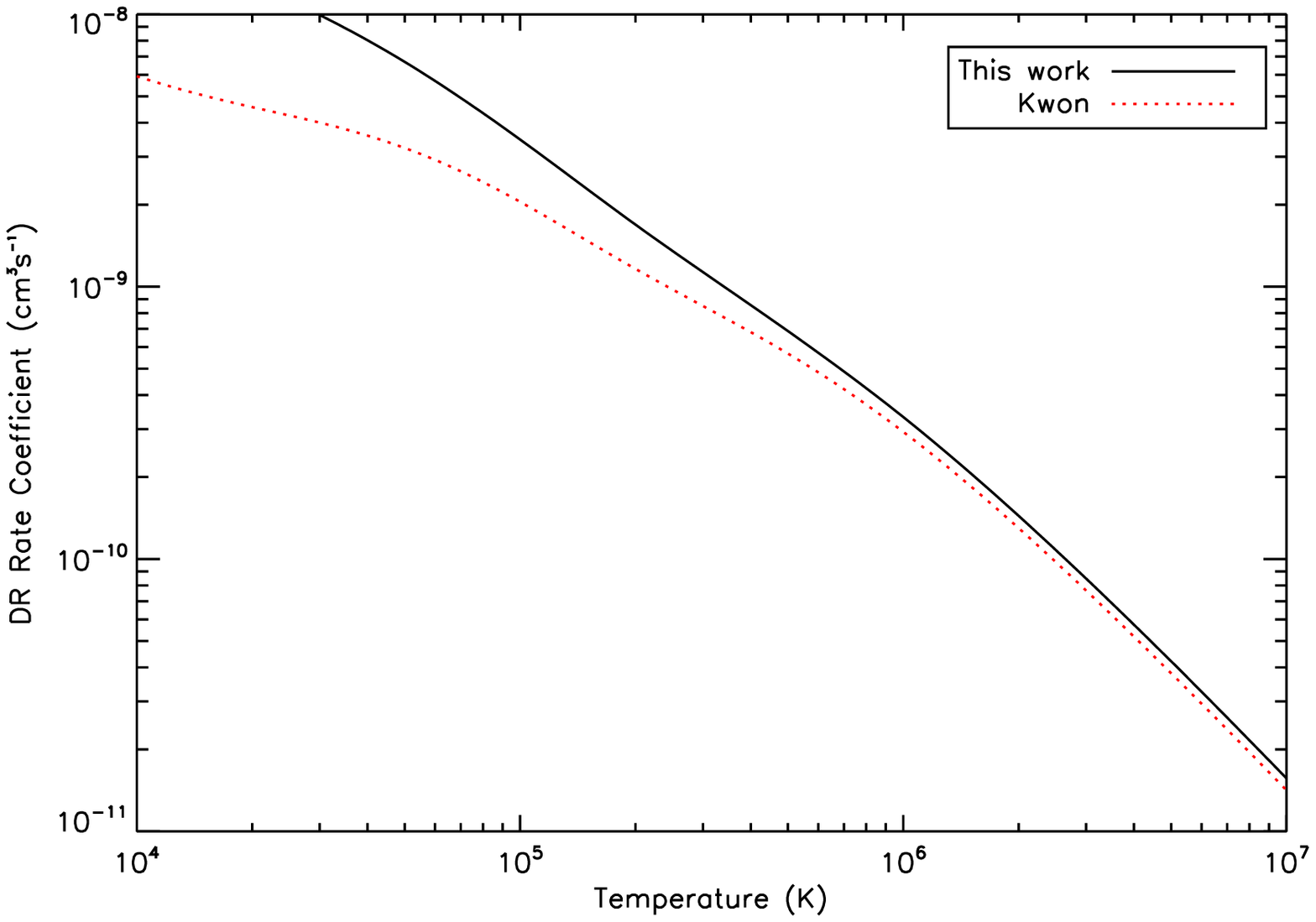}
\caption{Comparison between the total DR rate coefficients for 64-like W, 
calculated in this work (solid black), and Kwon \protect\cite{kwon2018a}
(red-dotted).}
\label{fig:w10comp}
\end{centering}
\end{figure}

In Figure \ref{fig:w10comp} we have plotted the total DR rate coefficient for 64-like
as calculated in the present work, and by Kwon \cite{kwon2018a}. At peak abundance temperature 
($\sim{7}\times{10}^{5}$K) we find our rates differ from Kwon's by 14\%. This 
difference decreases with increasing temperature, with our results differing by 10\% 
at $2\times{10}^{8}$K. For lower temperatures, our results and Kwon's diverge 
significantly, with the largest difference occuring at $1\times{10}^{4}$K of 68\%.

\subsection{65-like: W$^{ 9+}$}
\begin{figure}
\begin{centering}
\includegraphics[width=85mm]{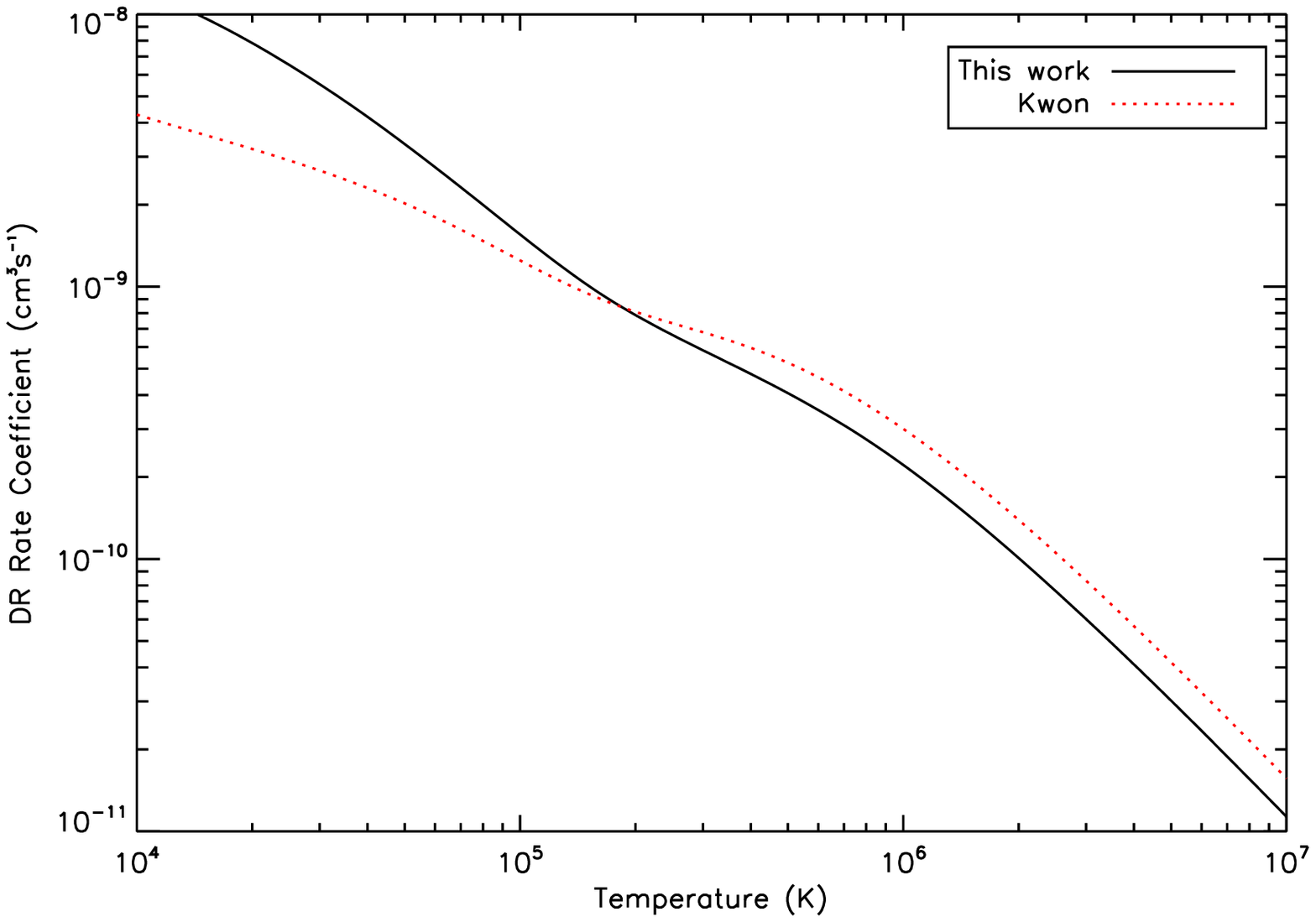}
\caption{Comparison between the total DR rate coefficients for 65-like W
calculated in this work (solid black), and Kwon \protect\cite{kwon2018a}
(red-dotted).}
\label{fig:w9comp}
\end{centering}
\end{figure}

We have plotted the total DR rate coefficients for 65-like as calculated in the 
present work in Figure \ref{fig:w9comp}, along with the result calculated by Kwon \cite{kwon2018a}. 
Reasonable agreement is seen at peak abundance temperature ($6.2\times{10}^{5}$K), 
with Kwon's result being $32\%$ larger than the present work. This difference increases
slightly with increasing temperature, reaching 38\% at $1.6\times{10}^{8}$K. Agreement
also deteriorates towards lower temperatures, with Kwon's DR rate coefficient being
$\sim{70}$\% smaller than the present work for temperatures $<1000$K.

\subsection{66-like: W$^{ 8+}$}
\begin{figure}
\begin{centering}
\includegraphics[width=85mm]{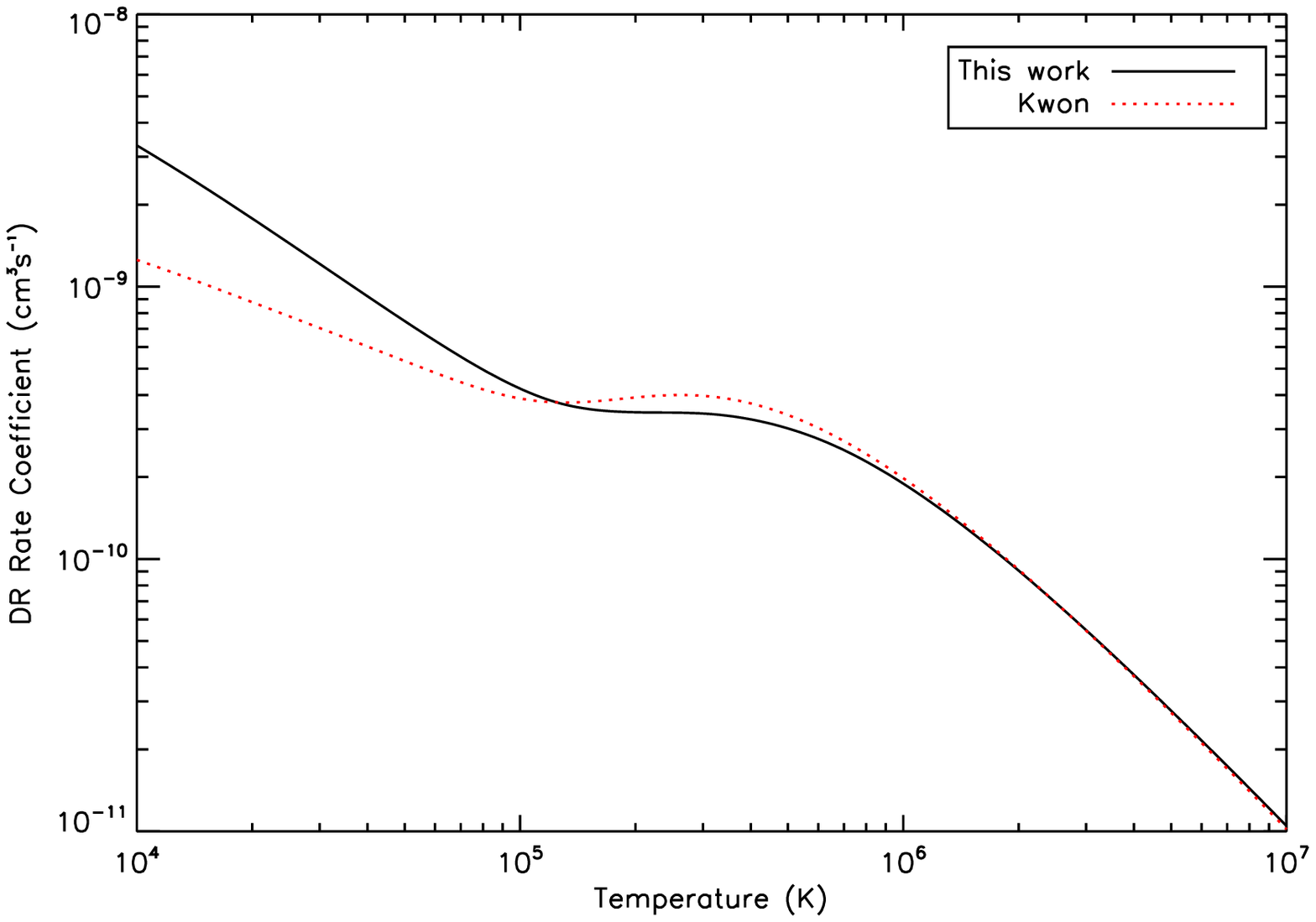}
\caption{Comparison between the total DR rate coefficients for 66-like W
calculated in this work (solid black), and Kwon \protect\cite{kwon2018a}
(red-dotted).}
\label{fig:w8comp}
\end{centering}
\end{figure}

In Figure \ref{fig:w8comp} we have plotted the total DR rate coeffcients as calculated
in the present work, and by Kwon \cite{kwon2018a}. Good agreement is seen at peak abundance 
temperature ($5.1\times{10}^{5}$K), with Kwon's DR rate coefficients being larger than
the present work by 11\%. Agreement improves to better than 10\% for temperatures
$>6\times{10}^{5}$K. The largest differences are seen at lower temperatures, being
$>80$\% for temperatures $<$ 1000K.

\subsection{67-like: W$^{ 7+}$}
\begin{figure}
\begin{centering}
\includegraphics[width=85mm]{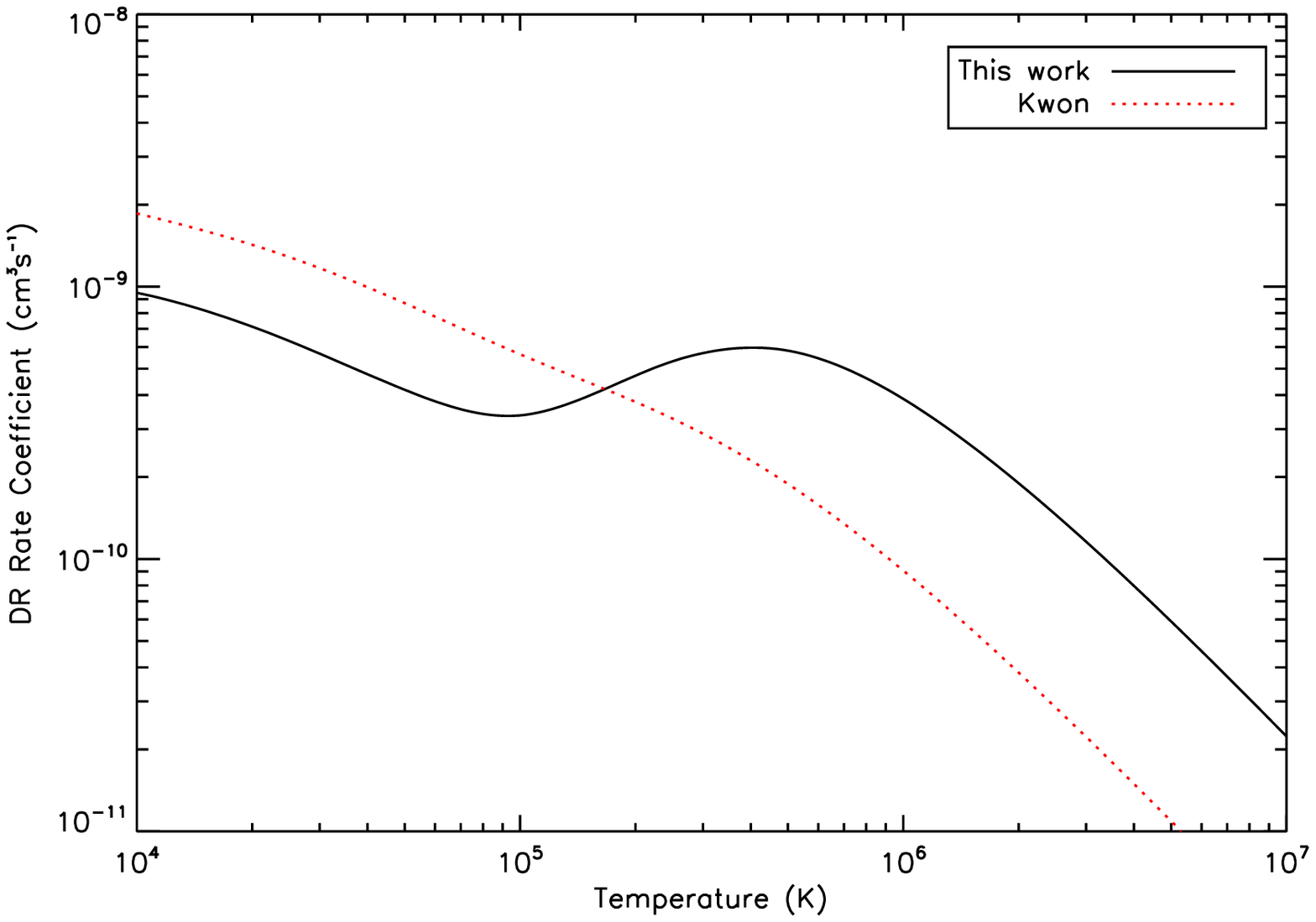}
\caption{Comparison between the total DR rate coefficients for 67-like W
calculated in this work (solid black), and Kwon \protect\cite{kwon2018a}
(red-dotted).}
\label{fig:w7comp}
\end{centering}
\end{figure}

In Figure \ref{fig:w7comp} we have plotted the total DR rate coefficients for 67-like
as calculated in the present work, and by Kwon \cite{kwon2018a}. Significant differences are seen
between both sets of DR rate coefficients across a wide range of temperatures. At 
peak abundance temperature ($4.2\times{10}^{5}$K), Kwon's DR rate coefficients are 
$\sim{60}$\% smaller than ours. With increasing temperature, the difference between 
our and Kwon's DR rate coefficients becomes constant, reaching $\sim{83}$\% at 
$\sim{1}\times{10}^{8}$K. Towards lower temperatures, the largest difference between
our data and Kwon's is seen at $\sim{1000}$K, where Kwon's DR rate coefficients are 
larger by a factor $\sim{3}$.

\subsection{68-like: W$^{ 6+}$}
\begin{figure}
\begin{centering}
\includegraphics[width=85mm]{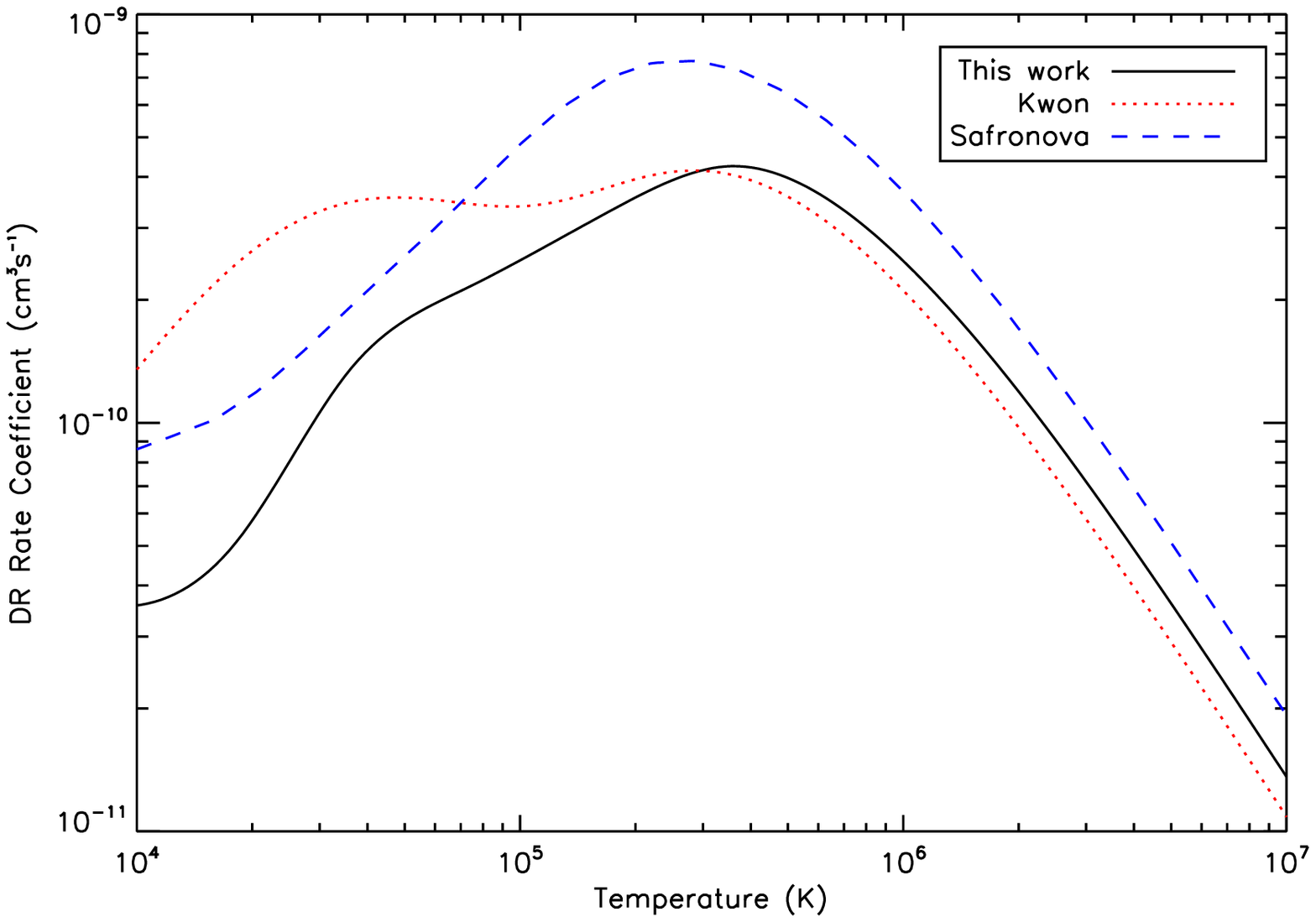}
\caption{Comparison between the total DR rate coefficients for 68-like W 
calculated in this work (solid black), Kwon \protect\cite{kwon2018a}
(red-dotted), and Safronova~\etal \protect\cite{usafronova2012b} 
(blue-dash).}
\label{fig:w6comp}
\end{centering}
\end{figure}

DR rate coefficients for this ion have been calculated by Kwon \cite{kwon2018a} and 
Safronova~\etal \cite{usafronova2012a,usafronova2012b}. In 
Figure \ref{fig:w6comp} we have plotted their results, along with those calculated
in this current work. Poor agreement is evident over a wide range of temperatures.
However, at peak abundance temperature ($2.7\times{10}^{5}$K), Kwon's DR rate 
coefficients are larger by $\sim{2}$\%, while Safronova~\etal's are larger by a factor
$\sim{2}$. Given the variation of the DR rate coefficient either side of the peak
abundance temperature, this agreement appears to be coincidental. Towards higher
temperatures ($>3\times{10}^{6}$K) the difference between our data and Kwon's remains
constant, with Kwon's DR rate coefficients being $\sim{20}$\% smaller than the present
calculation. In the case of Safronova's data for these temperatures, the difference 
varies from $\sim{40}-90$\% larger than the present values. Towards lower temperatures,
the difference between all datasets diverge strongly, with the largest differences
exceeding $\sim{9}$ dex at the lowest temperatures. 

\subsection{69-like: W$^{ 5+}$}
\begin{figure}
\begin{centering}
\includegraphics[width=85mm]{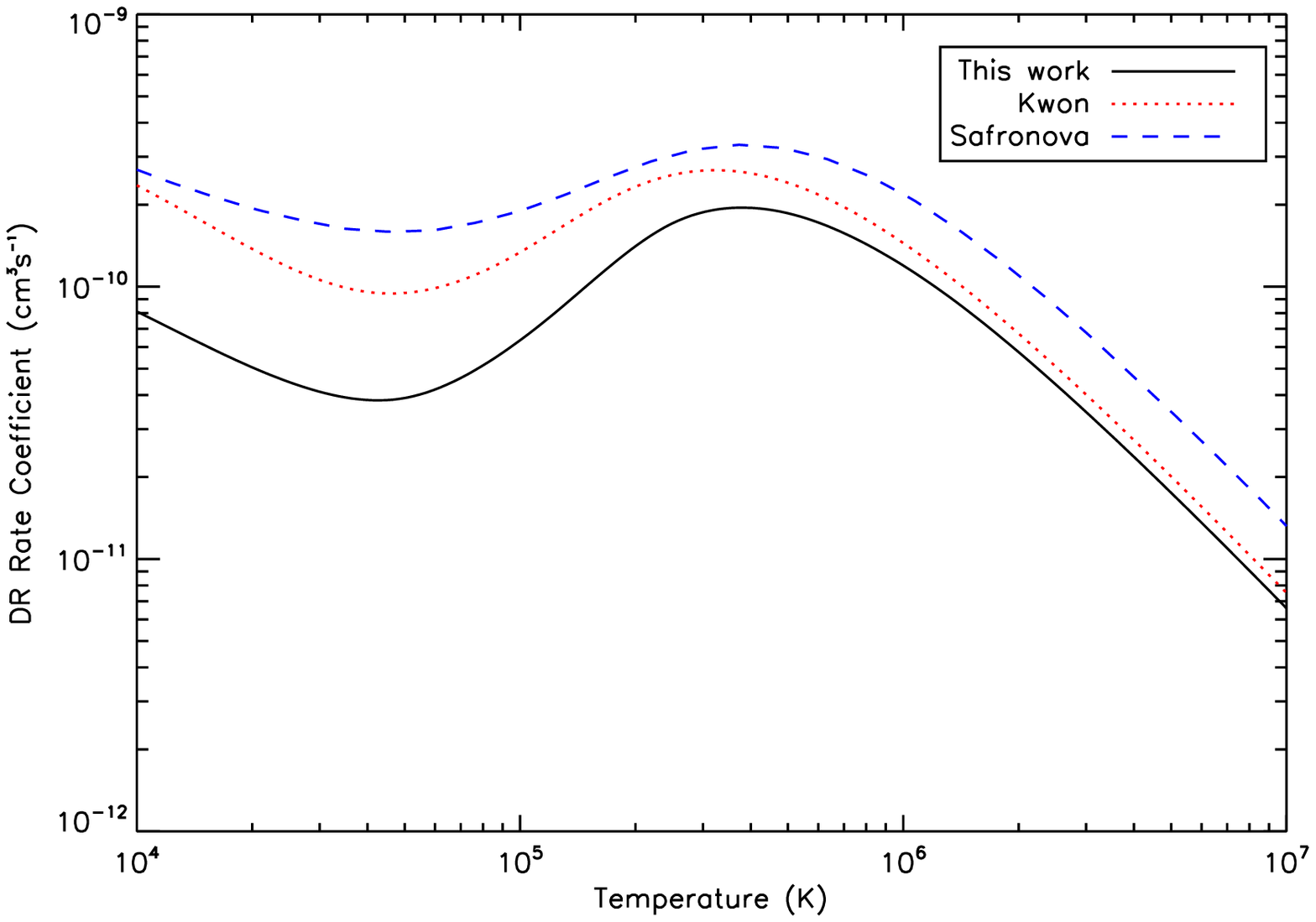}
\caption{Comparison between the total DR rate coefficients for 69-like W 
calculated in this work (solid black), Kwon \protect\cite{kwon2018a}
(red-dotted), and Safronova~\etal \protect\cite{usafronova2012a} 
(blue-dash).}
\label{fig:w5comp}
\end{centering}
\end{figure}

As with 68-like, DR rate coefficients have been calculated by Kwon \cite{kwon2018a} 
and Safronova~\etal \cite{usafronova2012a,usafronova2012b}. In
Figure \ref{fig:w5comp} we have plotted our results, along with those of Kwon and 
Safronova~\etal. The general trend of all three calculations appear to be in agreement.
At peak abundance temperature ($1.7\times{10}^{5}$K) the DR rate coefficients calculated
by Kwon and Safronova are larger than the present data by $\sim{80}$\% and a factor 
$\sim{2}$ respectively. Towards higher temperatures the agreement between the present
data and Kwon's results improves, with Kwon's DR rate coefficients being larger by
$\sim{13}$\%. This is not the case for Safronova~\etal's data, where the difference
varies between a factor 1.7-2.0 with increasing temperature. Agreement does not 
improve with decreasing temperature, with differences of a factor $\sim{9}$ and 
$\sim{4}$ for Kwon and Safronova~\etal respectively. 

\section{Ionization State Evolution}
In this section we consider the impact of our calculations in the context of ionization fractions.
We first consider the steady state ionisation fraction incorporating our data, and lastly, we
consider a time-dependent case.

\subsection{Steady state ionization}
\begin{figure*}
\begin{centering}
\includegraphics[width=170mm]{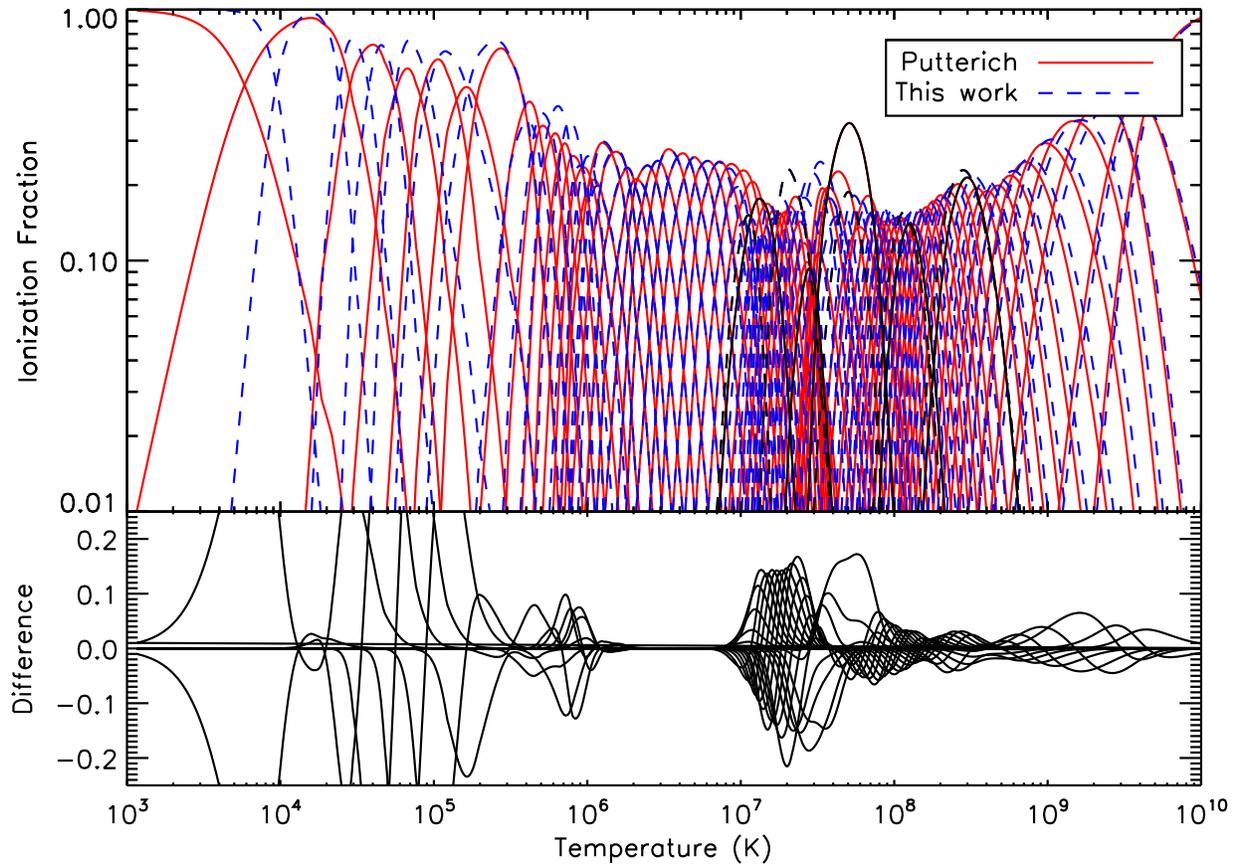}
\caption{Plot of the coronal, steady-state ionization fraction for tungsten. The red
fraction was calculated using the recombination rate coefficients of P\"{u}tterich~\etal
\protect\cite{putterich2008a}, and the ionization rate coefficients of Loch~\etal
\protect\cite{loch2005a}. The blue-dashed fraction was calculated using the recombination
rate coefficients calculated in this work (61- to 73-like), and the data from previous
work by Preval~\etal \cite{preval2016a,preval2017b,preval2018a} (00- to 46-like). We
used P\"{u}tterich~\etal's data for 47- to 60-like. From right to left, we have indicated the
positions of the closed-shell states 10-, 18-, 28-, 36-, and 46-like with black parabolas. 
The bottom plot is the arithmetic difference between the P\"{u}tterich~\etal fractions, and 
the present fraction.}
\label{fig:ionfrac}
\end{centering}
\end{figure*}

We now compare two sets of ionization fractions, calculated using recombination rate coefficient
data from this work (61- to 73-like) and \textit{The Tungsten Project} (00- to 46-like, 
\cite{preval2016a,preval2017b,preval2018a}), and using the scaled data from P\"{u}tterich~\etal 
\cite{putterich2008a}. For the ionization fraction calculated using the present data, we use
the data from P\"{u}tterich et al for 47- to 60-like, as we have not calculated data for these
charge states yet. In both cases, we use the ionization rate coefficients as calculated by 
Loch~\etal \cite{loch2005a}. The two ionization fractions are plotted in Figure \ref{fig:ionfrac},
along with the arithmetical difference between the two fractions. We have also indicated the position
of closed-shell charge states as a guide. In this plot, there is a large gap indicating zero 
difference between the two fractions. This is because the data for 47- to 60-like are the same 
in both fractions. Immediately obvious is the significant differences in the peak abundance temperature,
and the peak abundance fractions. In Table \ref{table:peaktemp} we have listed these peak fractions 
and temperatures for our ionization fraction, and that of P\"{u}tterich~\etal. We have also calculated 
the \% difference between the peak temperatures and fractions calculated in this work, and those 
from P\"{u}tterich~\etal. 

Looking at the plot overall, it is clear that the largest differences between our ionization
fraction and P\"{u}tterich~\etal's occurs for 61- to 73-like. This is indicative of the difference
in atomic structures used in both approaches, and also highlights the difficulty in calculating
a reliable atomic structure. As mentioned in Section \ref{authorcomp}, relatively good agreement between
our calculations and Kwon's \cite{kwon2018a} is seen for the higher ionization states. 
For the lower ionization states, this agreement deteriorates. Further calculations by other groups using different
codes could help to improve our understanding of the atomic structure for these ions.

\subsection{Time-dependent ionization}
\begin{figure*}
\begin{centering}
\includegraphics[width=170mm]{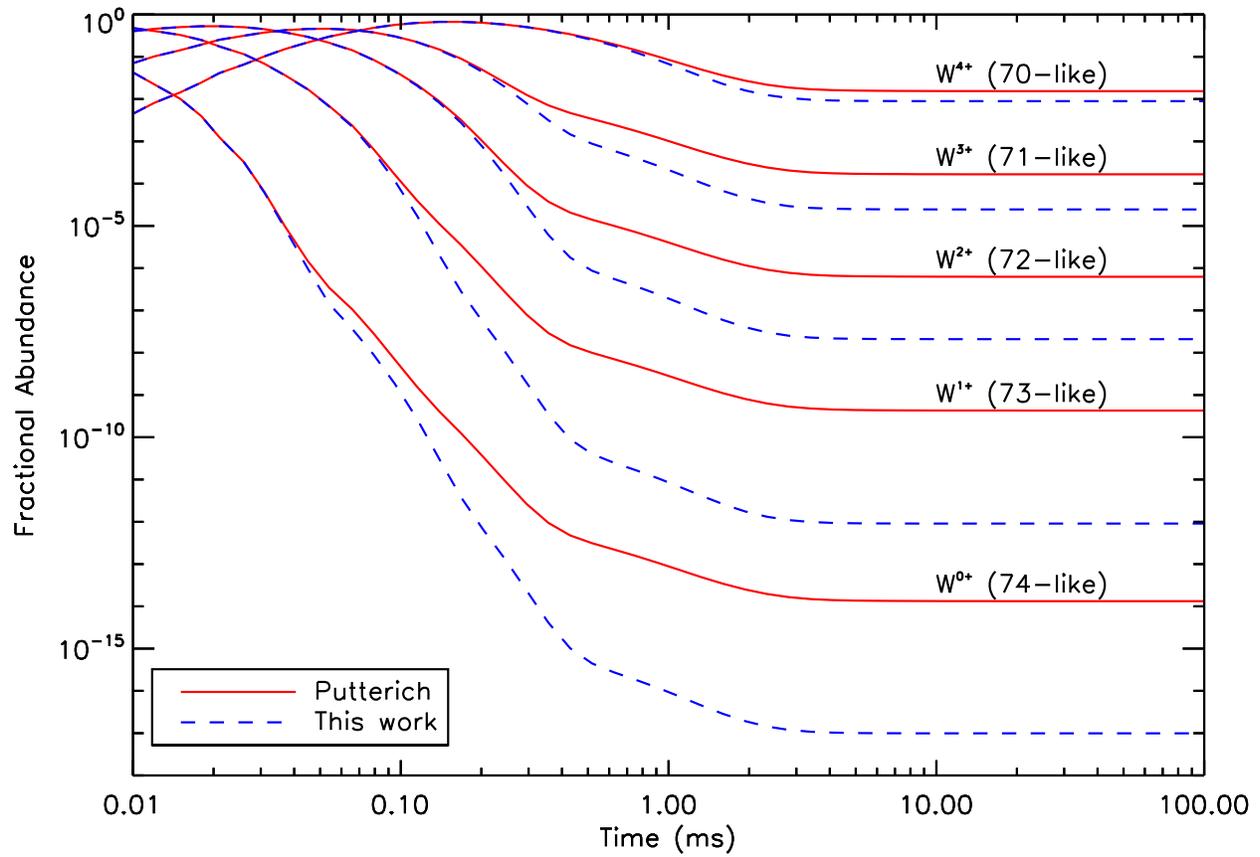}
\caption{Time evolution of the charge state distribution of 74- to 70-like tungsten 
for a 20eV plasma over 100ms. The red solid line is the case where the recombination rate 
coefficients of P\"{u}tterich~\etal \protect\cite{putterich2008a} were used, whereas 
the blue dashed line is the case where the present recombination rate coefficients were used.}
\label{fig:ionevol}
\end{centering}
\end{figure*}

To further illustrate the impact of our data, we considered the time evolution of the 
ionization fractions for a 20eV, fixed density plasma where a tungsten impurity was
introduced. This was done using the ADAS406 routine, the description for which can be 
found on the ADAS website\footnote{http://www.adas.ac.uk/}. In Figure \ref{fig:ionevol} 
we have plotted the evolution of five charge states spanning 70- to 74-like over a 
period of 100ms using the present data, and the recombination rate coefficients of 
P\"{u}tterich~\etal. While ionization is the dominant process, it can be clearly seen 
that the recombination rate coefficients determine the final equilibrium state. The 
differences seen can easily be attributed to the methods used in calculating the 
recombination rate coefficients.

\section{Conclusions}
We have presented a series of partial, final-state resolved DR and RR rate coefficients
for low-charge tungsten ions spanning W$^{13+}$ - W$^{1+}$. The present work constitutes
the first such DR/RR calculations performed for low ionization state tungsten. These data
will be paramount in modelling the collisional-radiative properties of the edge plasma
in magnetically confined, finite density plasmas such as those observed in JET and ITER.

We calculated an updated coronal, steady-state ionization balance for tungsten using all
of the recombination rate coefficient data calculated in \textit{The Tungsten Project},
and the ionization rate coefficients from Loch~\etal \cite{loch2005a}. We compared this 
with an ionization balance calculated using the recombination rate coefficients of 
P\"{u}tterich~\etal \cite{putterich2008a}, and found significant differences between 
the two fractions. In particular, there were large shifts in the peak fractions and 
temperatures. The majority of these large changes were for the lowest ionization states 
considered, illustrating the difficulty in calculating an accurate atomic structure for 
these ions.

We find our DR rate coefficients are in relatively good agreement with the few currently 
published, and we are able to reproduce the general trend of these data. This is in contrast
to the case of more highly-charged ions where substantial differences were found. However, large 
differences are seen towards lower temperatures. While this isn't a problem in the case
of highly charged states, in the case of singly, doubly, or triply ionized ions the
peak abundance temperature and peak fraction will be sensitive to threshold effects.
This was also evidenced by considering a time-dependent ionization fraction using the ADAS
routine ADAS406. It was shown that for a fixed density, 20 eV plasma with a tungsten 
impurity, the recombination rate coefficients used for 70- to 73-like had a significant 
impact on the final equilibrium state. Therefore, more work is required to constrain 
the DR rate coefficients further for these ionization states.

The main challenge in improving the DR rate coefficients for low ionization state tungsten
lies in constraining the positioning of near-threshold resonances. As well as extensive
theoretical work, experiment must also be used. Cryogenic storage ring experiments such as those
described in Spruck~\etal \cite{spruck2015a} and Von-Hahn~\etal \cite{vonhahn2016a} can
potentially be used in the case of low charge-state tungsten ions. As seen in Figure \ref{fig:ionevol},
significant differences were seen in the final equilibrium state of a 20 eV plasma with a tungsten impurity
when using the present data, or that of P\"{u}tterich~\etal \cite{putterich2008a}. The largest
differences between the two cases was seen for neutral-state tungsten. However, this difference
decreased with increasing residual charge, indicating a larger uncertainty in the DR rate coefficients
for the lowest charge states. Therefore, future experiments should focus on the near-neutral
charge states of tungsten.

The data presented in this paper, combined with our work on the $4d$-shell tungsten ions,
gives an indication of how the missing $4f$-shell DR rate coefficients will behave when they are
added. Currently, the open $4f$-shell problem in calculating DR rate coefficients is still
untenable by even the best computational systems without implementing some form of
statistical approximation. It is possible in the near future that a fully 
parallelised version of {\sc autostructure} could tackle this problem. The final paper in
\textit{The Tungsten Project} will consider the $4f$ shell as far as is possible with
{\sc autostructure}. We will then use a partitioning method as described in Badnell~\etal 
\cite{badnell2012a} to cover any ions we cannot compute directly.

\ack
SPP, NRB, and MGOM acknowledge the support of EPSRC grant EP/1021803
to the University of Strathclyde. All data calculated as part of this work are 
publicly available on the OPEN-ADAS website https://open.adas.ac.uk. SPP would like
to dedicate this paper to the memory of Mr Mark Preval, who passed away surrounded
by his family on 19th June 2018.

\section*{References}
\bibliography{simonpreval}

\providecommand{\newblock}{}
\begin{thebibliography}{10}
\expandafter\ifx\csname url\endcsname\relax
  \def\url#1{{\tt #1}}\fi
\expandafter\ifx\csname urlprefix\endcsname\relax\def\urlprefix{URL }\fi
\providecommand{\eprint}[2][]{\url{#2}}

\bibitem{post1977a}
Post D~E, Jensen R~V, Tarter C~B, Grasberger W~H and Lokke W~A 1977 {\em
  ADNDT\/} {\bf 20} 397--439
  \urlprefix\url{http://www.sciencedirect.com/science/article/pii/0092640X77900262}

\bibitem{post1995a}
Post D, Abdallah J, Clark R~E~H and Putvinskaya N 1995 {\em Phys. Plasmas\/}
  {\bf 2} 2328--2336
  \urlprefix\url{http://scitation.aip.org/content/aip/journal/pop/2/6/10.1063/1.871257}

\bibitem{putterich2008a}
P\"{u}tterich T, Neu R, Dux R, Whiteford A~D, O'Mullane M~G and the ASDEX
  Upgrade~Team 2008 {\em Plasma Phys. Control. Fusion\/} {\bf 50} 085016
  \urlprefix\url{http://stacks.iop.org/0741-3335/50/i=8/a=085016}

\bibitem{foster2008a}
Foster A~R 2008 {\em {On the Behaviour and Radiating Properties of Heavy
  Elements in Fusion Plasmas}\/} Ph.D. thesis University of Strathclyde
  http://www.adas.ac.uk/theses/foster\_thesis.pdf

\bibitem{burgess1965a}
{Burgess} A 1965 {\em ApJ\/} {\bf 141} 1588--1590
  \urlprefix\url{http://dx.doi.org/10.1086/148253}

\bibitem{badnell2003a}
Badnell N~R, O'Mullane M~G, Summers H~P, Altun Z, Bautista M~A, Colgan J,
  Gorczyca T~W, Mitnik D~M, Pindzola M~S and Zatsarinny O 2003 {\em Astronomy
  and Astrophysics\/} {\bf 406} 1151--1165 ISSN 0004-6361
  \urlprefix\url{http://dx.doi.org/10.1051/0004-6361:20030816}

\bibitem{chung2005a}
Chung H~K, Chen M~H, Morgan W~L, Ralchenko Y and Lee R~W 2005 {\em High Energy
  Density Physics\/} {\bf 1} 3--12
  \urlprefix\url{http://www.sciencedirect.com/science/article/pii/S1574181805000029}

\bibitem{badnell1986a}
Badnell N~R 1986 {\em J. Phys. B\/} {\bf 19} 3827
  \urlprefix\url{http://stacks.iop.org/0022-3700/19/i=22/a=023}

\bibitem{badnell1997a}
Badnell N~R 1997 {\em J. Phys. B\/} {\bf 30} 1
  \urlprefix\url{http://stacks.iop.org/0953-4075/30/i=1/a=005}

\bibitem{badnell2011a}
Badnell N~R 2011 {\em Comput. Phys. Commun.\/} {\bf 182} 1528
  \urlprefix\url{http://www.sciencedirect.com/science/article/pii/S0010465511001160}

\bibitem{preval2016a}
Preval S~P, Badnell N~R and O'Mullane M~G 2016 {\em Phys. Rev. A\/} {\bf 93}
  042703--999999
  \urlprefix\url{http://link.aps.org/doi/10.1103/PhysRevA.93.042703}

\bibitem{preval2017b}
Preval S~P, Badnell N~R and O'Mullane M~G 2017 {\em J. Phys. B\/} {\bf 50}
  105201 \urlprefix\url{http://stacks.iop.org/0953-4075/50/i=10/a=105201}

\bibitem{preval2018a}
Preval S~P, Badnell N~R and O'Mullane M~G 2018 {\em Journal of Physics B Atomic
  Molecular Physics\/} {\bf 51} 045004
  \urlprefix\url{http://stacks.iop.org/0953-4075/51/i=4/a=045004}

\bibitem{cowanbook1981}
Cowan R~D 1981 {\em {The Theory of Atomic Structure and Spectra}\/} Los Alamos
  Series in Basic and Applied Sciences (University of California Press) ISBN
  0520038215

\bibitem{barshalom2001a}
Bar-Shalom A, Klapisch M and Oreg J 2001 {\em JQSRT\/} {\bf 71} 169--188
  \urlprefix\url{http://www.sciencedirect.com/science/article/pii/S0022407301000668}

\bibitem{usafronova2012a}
Safronova U~I, Safronova A~S and Beiersdorfer P 2012 {\em J. Phys. B\/} {\bf
  45} 085001 \urlprefix\url{http://stacks.iop.org/0953-4075/45/i=8/a=085001}

\bibitem{usafronova2012b}
Safronova U~I and Safronova A~S 2012 {\em Phys. Rev. A\/} {\bf 85} 032507
  \urlprefix\url{http://link.aps.org/doi/10.1103/PhysRevA.85.032507}

\bibitem{usafronova2011a}
Safronova U~I, Safronova A~S, Beiersdorfer P and Johnson W~R 2011 {\em J. Phys.
  B\/} {\bf 44} 035005
  \urlprefix\url{http://stacks.iop.org/0953-4075/44/i=3/a=035005}

\bibitem{usafronova2016a}
Safronova U~I, Safronova A~S and Beiersdorfer P 2016 {\em J. Phys. B\/} {\bf
  49} 225002 \urlprefix\url{http://stacks.iop.org/0953-4075/49/i=22/a=225002}

\bibitem{usafronova2015a}
Safronova U~I, Safronova A~S and Beiersdorfer P 2015 {\em Phys. Rev. A\/} {\bf
  91} 062507 \urlprefix\url{http://dx.doi.org/10.1103/PhysRevA.91.062507}

\bibitem{usafronova2012c}
Safronova U~I, Safronova A~S and Beiersdorfer P 2012 {\em Phys. Rev. A\/} {\bf
  86} 042510 \urlprefix\url{http://link.aps.org/doi/10.1103/PhysRevA.86.042510}

\bibitem{usafronova2009a}
Safronova U~I, Safronova A~S and Beiersdorfer P 2009 {\em J. Phys. B\/} {\bf
  42} 165010 \urlprefix\url{http://stacks.iop.org/0953-4075/42/i=16/a=165010}

\bibitem{usafronova2009b}
Safronova U~I, Safronova A~S and Beiersdorfer P 2009 {\em ADNDT\/} {\bf 95}
  751--785
  \urlprefix\url{http://www.sciencedirect.com/science/article/pii/S0092640X09000278}

\bibitem{behar1997a}
Behar E, Peleg A, Doron R, Mandelbaum P and Schwob J~L 1997 {\em JQSRT\/} {\bf
  58} 449--469
  \urlprefix\url{http://www.sciencedirect.com/science/article/pii/S0022407397000526}

\bibitem{behar1999a}
Behar E, Mandelbaum P and Schwob J~L 1999 {\em Phys. Rev. A\/} {\bf 59}
  2787--999999 \urlprefix\url{http://link.aps.org/doi/10.1103/PhysRevA.59.2787}

\bibitem{behar1999b}
Behar E, Mandelbaum P and Schwob J~L 1999 {\em Eur. Phys. J. D\/} {\bf 7}
  157--161 \urlprefix\url{http://dx.doi.org/10.1007/s100530050361}

\bibitem{peleg1998a}
Peleg A, Behar E, Mandelbaum P and Schwob J~L 1998 {\em Phys. Rev. A\/} {\bf
  57} 3493--999999
  \urlprefix\url{http://link.aps.org/doi/10.1103/PhysRevA.57.3493}

\bibitem{gu2003a}
Gu M~F 2003 {\em ApJ\/} {\bf 590} 1131--999999
  \urlprefix\url{http://dx.doi.org/10.1086/375135}

\bibitem{li2012a}
Li B~W, O'Sullivan G, Fu Y~B and Dong C~Z 2012 {\em Phys. Rev. A\/} {\bf 85}
  052706 \urlprefix\url{http://link.aps.org/doi/10.1103/PhysRevA.85.052706}

\bibitem{li2014a}
Li M, Fu Y, Su M, Dong C and Koike F 2014 {\em Plasma Sci. Technol.\/} {\bf 16}
  182--187 \urlprefix\url{http://dx.doi.org/10.1088/1009-0630/16/3/02}

\bibitem{li2016a}
Li B, O'Sullivan G, Dong C and Chen X 2016 {\em J. Phys. B\/} {\bf 49}
  155201--999999
  \urlprefix\url{http://dx.doi.org/10.1088/0953-4075/49/15/155201}

\bibitem{meng2009a}
Meng F~C, Zhou L, Huang M, Chen C~Y, Wang Y~S and Zou Y~M 2009 {\em J. Phys.
  B\/} {\bf 42} 105203
  \urlprefix\url{http://stacks.iop.org/0953-4075/42/i=10/a=105203}

\bibitem{wu2015a}
Wu Z, Fu Y, Ma X, Li M, Xie L, Jiang J and Dong C 2015 {\em Atoms\/} {\bf 3}
  474 \urlprefix\url{http://www.mdpi.com/2218-2004/3/4/474}

\bibitem{kwon2018a}
Kwon D~H 2018 {\em Journal of Quantitative Spectroscopy and Radiative
  Transfer\/} {\bf 208} 64--70
  \urlprefix\url{http://www.sciencedirect.com/science/article/pii/S0022407317309263}

\bibitem{schippers2011a}
Schippers S, Bernhardt D, M\"{u}ller A, Krantz C, Grieser M, Repnow R, Wolf A,
  Lestinsky M, Hahn M, Novotn\'{y} O and Savin D~W 2011 {\em Phys. Rev. A\/}
  {\bf 83} 012711
  \urlprefix\url{https://link.aps.org/doi/10.1103/PhysRevA.83.012711}

\bibitem{spruck2014a}
Spruck K, Badnell N~R, Krantz C, Novotn\'{y} O, Becker A, Bernhardt D, Grieser
  M, Hahn M, Repnow R, Savin D~W, Wolf A, M\"{u}ller A and Schippers S 2014
  {\em Phys. Rev. A\/} {\bf 90} 032715
  \urlprefix\url{http://link.aps.org/doi/10.1103/PhysRevA.90.032715}

\bibitem{badnell2016a}
Badnell N~R, Spruck K, Krantz C, Novotn\'{y} O, Becker A, Bernhardt D, Grieser
  M, Hahn M, Repnow R, Savin D~W, Wolf A, M\"{u}ller A and Schippers S 2016
  {\em Phys. Rev. A\/} {\bf 93} 052703
  \urlprefix\url{http://link.aps.org/doi/10.1103/PhysRevA.93.052703}

\bibitem{badnell2012a}
Badnell N~R, Ballance C~P, Griffin D~C and O'Mullane M 2012 {\em Phys. Rev.
  A\/} {\bf 85} 052716
  \urlprefix\url{http://link.aps.org/doi/10.1103/PhysRevA.85.052716}

\bibitem{dzuba2012a}
Dzuba V~A, Flambaum V~V, Gribakin G~F and Harabati C 2012 {\em Phys. Rev. A\/}
  {\bf 86} 022714
  \urlprefix\url{https://link.aps.org/doi/10.1103/PhysRevA.86.022714}

\bibitem{dzuba2013a}
Dzuba V~A, Flambaum V~V, Gribakin G~F, Harabati C and Kozlov M~G 2013 {\em
  Phys. Rev. A\/} {\bf 88} 062713
  \urlprefix\url{https://link.aps.org/doi/10.1103/PhysRevA.88.062713}

\bibitem{berengut2015a}
Berengut J~C, Harabati C, Dzuba V~A, Flambaum V~V and Gribakin G~F 2015 {\em
  Phys. Rev. A\/} {\bf 92} 062717
  \urlprefix\url{https://link.aps.org/doi/10.1103/PhysRevA.92.062717}

\bibitem{harabati2017a}
Harabati C, Berengut J~C, Flambaum V~V and Dzuba V~A 2017 {\em Journal of
  Physics B: Atomic, Molecular and Optical Physics\/} {\bf 50} 125004
  \urlprefix\url{http://stacks.iop.org/0953-4075/50/i=12/a=125004}

\bibitem{demura2017a}
Demura A~V, Leont'iev D~S, Lisitsa V~S and Shurygin V~A 2017 {\em Journal of
  Experimental and Theoretical Physics\/} {\bf 125} 663--678
  \urlprefix\url{https://doi.org/10.1134/S1063776117090138}

\bibitem{krantz2017a}
Krantz C, Badnell N~R, M\"{u}ller A, Schippers S and Wolf A 2017 {\em Journal
  of Physics B: Atomic, Molecular and Optical Physics\/} {\bf 50} 052001
  \urlprefix\url{http://stacks.iop.org/0953-4075/50/i=5/a=052001}

\bibitem{kwon2017a}
Kwon D~H, Lee W, Preval S, Ballance C~P, Behar E, Colgan J, Fontes C~J, Nakano
  T, Li B, Ding X, Dong C~Z, Fu Y~B, Badnell N~R, O'Mullane M, Chung H~K and
  Braams B~J 2017 {\em Atomic Data and Nuclear Data Tables\/}
  \urlprefix\url{http://www.sciencedirect.com/science/article/pii/S0092640X17300220}

\bibitem{huang2018a}
Huang Z~K, Wen W~Q, Xu X, Mahmood S, Wang S~X, Wang H~B, Dou L~J, Khan N,
  Badnell N~R, Preval S~P, Schippers S, Xu T~H, Yang Y, Yao K, Xu W~Q, Chuai
  X~Y, Zhu X~L, Zhao D~M, Mao L~J, Ma X~M, Li J, Mao R~S, Yuan Y~J, Wu B, Sheng
  L~N, Yang J~C, Xu H~S, Zhu L~F and Ma X 2018 {\em ApJS\/} {\bf 235} 2
  \urlprefix\url{http://dx.doi.org/10.3847/1538-4365/aaa5b3}

\bibitem{synge1957a}
Synge J~L 1957 {\em {The relativistic gas}\/} Series in physics (North-Holland
  Pub. Co.) \urlprefix\url{https://books.google.co.uk/books?id=HM1-AAAAIAAJ}

\bibitem{loch2005a}
Loch S~D, Ludlow J~A, Pindzola M~S, Whiteford A~D and Griffin D~C 2005 {\em
  Phys. Rev. A\/} {\bf 72} 052716--999999
  \urlprefix\url{http://link.aps.org/doi/10.1103/PhysRevA.72.052716}

\bibitem{NIST_ASD2}
Kramida A, Yu, Reader J,  and Team N~A 2018 {NIST Atomic Spectra Database} NIST
  Atomic Spectra Database (ver. 5.5.6), [Online]. Available:
  https://physics.nist.gov/asd [2018, November 5]. National Institute of
  Standards and Technology, Gaithersburg, MD.

\bibitem{kramida2006a}
Kramida A~E and Shirai T 2006 {\em Journal of Physical and Chemical Reference
  Data\/} {\bf 35} 423--683 (\textit{Preprint}
  \eprint{https://doi.org/10.1063/1.1836763})
  \urlprefix\url{https://doi.org/10.1063/1.1836763}

\bibitem{kramida2006b}
Kramida A~E and Reader J 2006 {\em Atomic Data and Nuclear Data Tables\/} {\bf
  92} 457--479
  \urlprefix\url{http://www.sciencedirect.com/science/article/pii/S0092640X06000167}

\bibitem{spruck2015a}
Spruck K, Becker A, Fellenberger F, Grieser M, von Hahn R, Klinkhamer V,
  Novotn\'{y} O, Schippers S, Vogel S, Wolf A and Krantz C 2015 {\em Rev. Sci.
  Inst.\/} {\bf 86} 023303 \urlprefix\url{https://doi.org/10.1063/1.4907352}

\bibitem{vonhahn2016a}
von Hahn R, Becker A, Berg F, Blaum K, Breitenfeldt C, Fadil H, Fellenberger F,
  Froese M, George S, G\"{o}ck J, Grieser M, Grussie F, Guerin E~A, Heber O,
  Herwig P, Karthein J, Krantz C, Kreckel H, Lange M, Laux F, Lohmann S, Menk
  S, Meyer C, Mishra P~M, Novotn\'{y} O, O'Connor A~P, Orlov D~A, Rappaport
  M~L, Repnow R, Saurabh S, Schippers S, Schr\"{o}ter C~D, Schwalm D,
  Schweikhard L, Sieber T, Shornikov A, Spruck K, Sunil~Kumar S, Ullrich J,
  Urbain X, Vogel S, Wilhelm P, Wolf A and Zajfman D 2016 {\em Rev. Sci.
  Inst.\/} {\bf 87} 063115 \urlprefix\url{https://doi.org/10.1063/1.4953888}

\end{thebibliography}
\newpage

\clearpage

\begin{table*}
\renewcommand{\arraystretch}{1.2}
\caption{List of core $N$-electron configurations included in our calculations for each 
charge state. The $(N+1)$-electron configurations were obtained by adding an additional
electron to all of the $N$-electron configurations. We also indicate the core excitations 
extracted from these calculations.}
\begin{tabular}{@{}lll}
\hline
\hline
Ion-like & Core excitations & N-electron configurations \\
\hline
61-like & 4--4, 4--5, 4--6, 5--4, & $4d^{10}4f^{13}5snl$, $n=4-6$, $\ell=0-5$ \\
        & 5--5, 5--6              & $4d^{10}4f^{12}5s^{2}nl$, $n=5-6$, $\ell=0-5$ \\
        &                         & $4d^{9}4f^{13}5s^{2}nl$, $n=4-6$, $\ell=0-5$ \\
        &                         & $4d^{10}4f^{13}6s^{2}$ \\
        &                         & $4d^{10}4f^{13}6s6p$ \\
        &                         & $4d^{10}4f^{13}6p^{2}$ \\
\hline
62-like & 4--5, 4--6, 5--5, 5--6  & $4d^{10}4f^{14}5snl$, $n=5-6$, $\ell=0-5$ \\
        &                         & $4d^{10}4f^{13}5s^{2}nl$, $n=5-6$, $\ell=0-5$ \\
        &                         & $4d^{9}4f^{14}5s^{2}nl$, $n=5-6$, $\ell=0-5$ \\
        &                         & $4d^{10}4f^{14}6s^{2}$ \\
        &                         & $4d^{10}4f^{14}6s6p$ \\
        &                         & $4d^{10}4f^{14}6p^{2}$ \\
\hline
63-like & 4--4, 4--5, 4--6, 4--7, & $4d^{10}4f^{13}5s^{2} 5pnl$, $n=4-7$, $\ell=0-6$ \\
        & 5--4, 5--5, 5--6, 5--7  & $4d^{10}4f^{13}5s5p^{2}nl$, $n=4-7$, $\ell=0-6$ \\
        &                         & $4d^{10}4f^{12}5s^{2}5p^{2}nl$, $n=4-7$, $\ell=0-6$ \\
        &                         & $4d^{9}4f^{13}5s^{2}5p^{2}nl$, $n=4-7$, $\ell=0-6$ \\
\hline
64-like & 4--4, 4--5, 4--6, 5--5, & $4f^{14}5s^{2}5pnl$, $n=5-6$, $\ell=0-5$ \\
        & 5--6                    & $4f^{14}5s5p^{2}nl$, $n=5-6$, $\ell=0-5$ \\
        &                         & $4f^{13}5s^{2}5p^{2}nl$, $n=4-6$, $\ell=0-5$ \\
        &                         & $4f^{14}5s^{2}6s^{2}$ \\
        &                         & $4f^{14}5s^{2}6s6p$ \\
        &                         & $4f^{14}5s^{2}6p^{2}$ \\
\hline
65-like & 4--4, 4--5, 4--6, 5--5, & $4f^{14}5s^{2}5p^{2}nl$, $n=5-6$, $\ell=0-5$ \\
        & 5--6                    & $4f^{14}5s5p^{3}nl$, $n=5-6$, $\ell=0-5$ \\
        &                         & $4f^{13}5s^{2}5p^{3}nl$, $n=4-6$, $\ell=0-5$ \\
        &                         & $4f^{14}5s^{2}5p6s^{2}$ \\
        &                         & $4f^{14}5s^{2}5p6s6p$ \\
        &                         & $4f^{14}5s^{2}5p6p^{2}$ \\
\hline
66-like & 4--4, 4--5, 4--6, 5--5, & $4f^{14}5s^{2}5p^{3}nl$, $n=5-6$, $\ell=0-5$ \\
        & 5--6                    & $4f^{14}5s5p^{4}nl$, $n=5-6$, $\ell=0-5$ \\
        &                         & $4f^{13}5s^{2}5p^{4}nl$, $n=4-6$, $\ell=0-5$ \\
        &                         & $4f^{14}5s^{2}5p^{2}6s^{2}$ \\
        &                         & $4f^{14}5s^{2}5p^{2}6s6p$ \\
        &                         & $4f^{14}5s^{2}5p^{2}6p^{2}$ \\
\hline
67-like & 4--4, 4--5, 4--6, 5--4, & $4f^{13}5s^{2}5p^{5}nl$, $n=4-6$, $\ell=0-5$ \\
        & 5--5, 5--6              & $4f^{12}5s^{2}5p^{6}nl$, $n=5-6$, $\ell=0-5$ \\
        &                         & $4f^{13}5s^{2}5p^{4}6s^{2}$ \\
        &                         & $4f^{13}5s^{2}5p^{4}6s6p$ \\
        &                         & $4f^{13}5s^{2}5p^{4}6p^{2}$ \\
\hline
\hline
\label{table:coreconfig}
\end{tabular}
\renewcommand{\arraystretch}{1.0}
\end{table*}

\begin{table*}
\renewcommand{\arraystretch}{1.2}
\contcaption{Continued.}
\begin{tabular}{@{}lll}
\hline
\hline
Ion-like & Core excitations & N-electron configurations \\
\hline
68-like & 4--5, 4--6, 5--5, 5--6  & $4f^{14}5s^{2}5p^{5}nl$, $n=5-6$, $\ell=0-5$ \\
        &                         & $4f^{13}5s^{2}5p^{6}nl$, $n=4-6$, $\ell=0-5$ \\
        &                         & $4f^{14}5s^{2}5p^{4}6s^{2}$ \\
        &                         & $4f^{14}5s^{2}5p^{4}6s6p$ \\
        &                         & $4f^{14}5s^{2}5p^{4}6p^{2}$ \\
\hline
69-like & 4--4, 4--5, 4--6, 5--4, & $4f^{14}5s^{2}5p^{6}nl$, $n=5-6$, $\ell=0-5$ \\
        & 5--5, 5--6              & $4f^{14}5s^{2}5p^{5}5dnl$, $n=5-6$, $\ell=0-5$ \\
        &                         & $4f^{13}5s^{2}5p^{6}5dnl$, $n=4-6$, $\ell=0-5$ \\
        &                         & $4f^{14}5s^{2}5p^{5}6s^{2}$ \\
        &                         & $4f^{14}5s^{2}5p^{5}6s6p$ \\
        &                         & $4f^{14}5s^{2}5p^{5}6p^{2}$ \\
\hline
70-like & 5--5, 5--6              & $5p^{6}5dnl$, $n=5-6$, $\ell=0-5$ \\
        &                         & $5p^{5}5d^{2}nl$, $n=5-6$, $\ell=0-5$ \\
        &                         & $5p^{6}6s^{2}$ \\
        &                         & $5p^{6}6s6p$ \\
        &                         & $5p^{6}6p^{2}$ \\
\hline
71-like & 4--5, 4--6, 5--5, 5--6  & $4f^{14}5s^{2}5p^{6}5d^{2}nl$, $n=4-6$, $\ell=0,5$ \\
        &                         & $4f^{14}5s^{2}5p^{5}5d^{3}nl$, $n=4-6$, $\ell=0,5$ \\
        &                         & $4f^{14}5s5p^{6}5d^{3}nl$, $n=4-6$, $\ell=0,5$ \\
        &                         & $4f^{13}5s^{2}5p^{6}5d^{3}nl$, $n=4-6$, $\ell=0,5$ \\
\hline
72-like & 5--5, 5--6              & $5d^{3}nl$, $n=5-6$, $\ell=0-5$ \\
        &                         & $5d^{2}6s^{2}$ \\
        &                         & $5d^{2}6s6p$ \\
        &                         & $5d^{2}6p^{2}$ \\
\hline
73-like & 5--5, 5--6              & $5d^{4}nl$, $n=5-6$, $\ell=0-5$ \\
        &                         & $5d^{3}6snl$, $n=5-6$, $\ell=0-5$ \\
\hline
\hline
\label{table:coreconfig}
\end{tabular}
\renewcommand{\arraystretch}{1.0}
\end{table*}

\begin{table}
\caption{List of scaling parameters $\lambda_{n\ell}$ employed for each ionization 
state.}
\begin{tabular}{@{}lccccccc}
\hline
\hline
Ion-like & Symbol & $\lambda_{n\ell}$ \\
\hline
61-like & W$^{13+}$ & 0.99 \\
62-like & W$^{12+}$ & 0.98 \\
63-like & W$^{11+}$ & 1.00 \\
64-like & W$^{10+}$ & 0.99 \\
65-like & W$^{ 9+}$ & 0.99 \\
66-like & W$^{ 8+}$ & 0.99 \\
67-like & W$^{ 7+}$ & 0.98 \\
68-like & W$^{ 6+}$ & 0.98 \\
69-like & W$^{ 5+}$ & 0.97 \\
70-like & W$^{ 4+}$ & 0.96 \\
71-like & W$^{ 3+}$ & 0.96 \\
72-like & W$^{ 2+}$ & 0.96 \\
73-like & W$^{ 1+}$ & 0.96 \\
\hline
\hline
\label{table:params}
\end{tabular}
\end{table}

\begin{table*}
\caption{Comparison of peak abundance temperatures and fractions as calculated using 
P\"{u}tterich~\etal$\!\!$'s data \cite{putterich2008a}, and P\"{u}tterich~\etal$\!\!$'s 
data with 01- to 46-like, and 61- to 74-like replaced with our data. The ionization 
rate coefficients originate from Loch~\etal \cite{loch2005a}. Note $[x]=10^{x}$.}
\begin{tabular}{@{}lccccccc}
\hline
\hline
Ion-like & Charge & Putt $T_{\mathrm{peak}}$ & Putt $f_{\mathrm{peak}}$ & This work $T_{\mathrm{peak}}$ & This work $f_{\mathrm{peak}}$ & $\Delta{T}$\% & $\Delta{f}$\% \\
\hline
61-like & W$^{13+}$ & 1.10[+6] & 0.210 & 1.16[+6] & 0.199 &  5.26 & -5.13 \\
62-like & W$^{12+}$ & 9.45[+5] & 0.236 & 1.01[+6] & 0.261 &  6.68 &  10.6 \\
63-like & W$^{11+}$ & 8.08[+5] & 0.261 & 8.96[+5] & 0.239 &  10.9 & -8.29 \\
64-like & W$^{10+}$ & 7.14[+5] & 0.293 & 7.72[+5] & 0.364 &  8.10 &  24.1 \\
65-like & W$^{ 9+}$ & 6.18[+5] & 0.322 & 6.44[+5] & 0.413 &  4.14 &  28.3 \\
66-like & W$^{ 8+}$ & 5.14[+5] & 0.345 & 5.06[+5] & 0.387 & -1.46 &  12.2 \\
67-like & W$^{ 7+}$ & 4.20[+5] & 0.430 & 4.08[+5] & 0.360 & -2.74 & -16.3 \\
68-like & W$^{ 6+}$ & 2.74[+5] & 0.701 & 2.54[+5] & 0.751 & -7.09 &  7.15 \\
69-like & W$^{ 5+}$ & 1.63[+5] & 0.492 & 1.19[+5] & 0.683 & -27.3 &  38.9 \\
70-like & W$^{ 4+}$ & 1.06[+5] & 0.634 & 6.95[+4] & 0.758 & -34.6 &  19.7 \\
71-like & W$^{ 3+}$ & 6.74[+4] & 0.583 & 4.55[+4] & 0.722 & -32.5 &  23.9 \\
72-like & W$^{ 2+}$ & 4.02[+4] & 0.725 & 2.92[+4] & 0.762 & -27.4 &  5.22 \\
73-like & W$^{ 1+}$ & 1.60[+4] & 0.925 & 1.62[+4] & 0.964 &  1.13 &  4.21 \\
74-like & W$^{ 0+}$ & 1.16[+3] & 1.000 & 4.71[+3] & 1.000 &  306. &  0.00 \\
\hline
\hline
\label{table:peaktemp}
\end{tabular}
\end{table*}


\end{document}